\documentclass{aa}  
\usepackage{keyval} 
\usepackage{booktabs}
\usepackage{xcolor}
\usepackage{color}
\usepackage{graphicx}
\usepackage{txfonts}
\usepackage{lipsum}
\usepackage{subcaption}   
\usepackage{setspace}
\usepackage{ulem}
\usepackage{lscape}             
\usepackage{placeins}           
\usepackage{natbib}                                

\newcommand{\um}{\textmu m}

\newcommand{\IH}{I_{\textit{Herschel}}}
\newcommand{\IS}{I_{\textit{SOFIA}}}
\newcommand{\sofia}{\textit{SOFIA}}
\newcommand{\hawc}{\textit{HAWC+}}

\begin{document}


 \title{SIMPLIFI -- Study of Interstellar Magnetic Polarization: A Legacy Investigation of Filaments. II. \\Enhancement of grain alignment near embedded protostars in the DR21 Ridge}
\titlerunning{SIMPLIFI II: Grain alignment near protostars in DR21}


   \author{S. Kumar\inst{1}
        \and T.G.S. Pillai\inst{2} 
        \and G. V. Panopoulou \inst{1}
        \and J. Kauffmann \inst{2}
        \and L. N. Tram \inst{3}
        \and S. Reissl \inst{4}
        \and D. Clemens\inst{5}
        \and V. J. M. Le Gouellec\inst{6}
        \and M. Heyer\inst{7}
        \and L. M. Fissel\inst{8}
        \and P. F. Goldsmith\inst{9}
        \and D. Seifried \inst{10}
        \and G. A. P. Franco \inst{11}
        }

   \institute{Department of Space, Earth and Environment, Chalmers University of Technology, 412 96, Gothenburg, Sweden \email{siddharth.kumar@chalmers.se}
            \and Haystack Observatory, Massachusetts Institute of Technology, 99 Millstone Rd., Westford, MA 01886, USA
            \and Leiden Observatory, Leiden University, PO Box 9513, 2300 RA Leiden, The Netherlands
            \and Zentrum f$\mathrm{\ddot{u}}$r Astronomie der Universität Heidelberg, Institut für Theoretische Astrophysik, Albert-Ueberle-Str. 2, 69120 Heidelberg, Germany
            \and Institute for Astrophysical Research, Boston University, 725 Commonwealth Avenue, Boston, MA 02215, USA
           \and NASA Ames Research Center, Space Science and Astrobiology Division, M.S. 245-6 Moffett Field, CA 94035, USA
           \and Department of Astronomy, University of Massachusetts, Amherst, MA 01003, USA
           \and Department of Physics, Engineering Physics and Astronomy, Queen’s University, 64 Bader Lane, Kingston, ON, Canada
           \and Jet Propulsion Laboratory, California Institute of Technology, 4800 Oak Grove Drive, Pasadena, CA 91109, USA
           \and Universität zu Köln, I. Physikalisches Institut, Z$\mathrm{\ddot{u}}$lpicher Str. 77, D-50937 Köln, Germany
           \and Departamento de Física -- ICEx -- UFMG Caixa Postal 702, 30.123-970 Belo Horizonte, Brazil}

   \date{Received September 30, 20XX}

 
  \abstract
  {Thermal dust continuum polarimetry is a powerful indirect probe of magnetic field geometry in dense molecular clouds while at the same time providing information on the alignment of dust grains with the magnetic field. The leading theory of grain alignment, Radiative Torque Alignment (RAT), has been successful in explaining a variety of observations, including the loss of polarization fraction toward high column densities. One prediction of RAT is that an increase in grain alignment efficiency should be observed in the environments surrounding protostars, due to radiation from the embedded source. However, observational confirmation of this prediction remains scarce.


}
   {In this study, we sought to test the theoretical prediction of enhanced grain alignment near protostars in the high-mass star forming region DR21 using 214 $\mu m$ \textit{SOFIA/HAWC+} observations.}
   {We investigated the correlation of the polarization fraction of dust emission, $p$, and the polarization angular dispersion, $S$, with respect to total intensity. We also probed intrinsic dust polarization properties using the product $S\times p$ as a proxy.
   }
   {We detected significant polarization fractions even at the highest intensities, where strong depolarization is typically expected. The polarization fraction-intensity trend flattens at $I > 1.6^{+0.3}_{-0.3} \times 10^4$ MJy/sr (\(N_{\rm H_2}\) $\sim  2\times 10^{23}$ $\rm{cm}^{-2}$). We compared the observed trends with predictions from an analytical model of a centrally heated envelope surrounding an embedded luminous protostar. The predictions from the simple model agree well with the observed trends. Our results provide strong support for enhancement of grain alignment by local radiation from embedded sources.}
   {}

\maketitle

\section{Introduction}

The discovery of starlight polarization \citep{1949Hiltner,1949Sci...109..166H} provided indirect evidence for the existence of large-scale interstellar magnetic fields. Subsequent studies demonstrated that this polarization is caused by the dichroic extinction of starlight by dust grains aligned with the magnetic fields \citep{1949Sci...109..165H}. Since then, several theories have been introduced to explain the mechanism of dust alignment with the magnetic field \citep[see][for a review]{2015ARA&A..53..501A}. \cite{1951ApJ...114..206D} proposed that paramagnetic dissipation of rotational kinetic energy can result in alignment of the angular momentum of dust grains with the magnetic field. They proposed that elongated dust grains rotate around their shorter axis, that is, along their axis of maximum moment of inertia, and subsequently paramagnetic dissipation of rotational kinetic energy aligns the spinning grains' angular momentum with the magnetic field. Their proposed mechanism (now known as the DG mechanism) correctly predicts the orientation of dust grains in the presence of magnetic fields but fails other tests, such as predicting better alignment for larger grains. The alignment of grains by mechanical torques exerted by gas flows was proposed as a complementary mechanism to paramagnetic relaxation \citep{1952Natur.169..322G,1952MNRAS.112..215G,2007ApJ...669L..77L,2018ApJ...852..129H} and modeled in \cite{2023A&A...674A..47R}. However, this mechanism was unable to predict increased alignment from $\rm H_2$ formation and the loss of alignment at $A_V \sim20$ mag \citep[see][for a review]{2015ARA&A..53..501A}.

The current leading theoretical framework for dust grain alignment is the Radiative Torque (RAT) alignment mechanism \citep{1976Ap&SS..43..291D,1996AAS...189.1602D,2007ApJ...669L..77L,2018ApJ...852..129H}. It has been successful in explaining the dependence of polarization fraction on intensity toward star-forming regions \citep{2022FrASS...9.3927T}. It posits that alignment happens in two steps: (1) irregularly shaped dust grains exposed to anisotropic radiation fields experience torques that spin them up to supra-thermal rotation; (2) paramagnetic dust grains in supra-thermal rotation align their angular momentum with the magnetic field via the Barnett effect  (magnetization of a rotating body due to coupling of spin with rotation) \citep{Barnett_1909,Purcell_1979}. Several observational studies have confirmed key predictions of RAT theory such as the dependence of alignment on the relative angle between the radiation field anisotropy and the magnetic field direction \citep{2010ApJ...720.1045A} and the dependence of polarization with dust temperature \citep{2021ApJ...923..130T} supporting its relevance in a range of astrophysical conditions. However, recent modeling of RATs within MHD simulations calls into question some of these observational tests. \citet{2020A&A...640A.118R} showed that alignment based on RAT and perfectly aligned dust grains yielded similar polarization maps in diffuse and translucent ISM conditions, and that grain alignment can only be distinguished in denser environments ($N_{H_2} > 10^{22} \mathrm{cm}^{-2}$).

Star-forming regions serve as excellent astronomical laboratories for observational testing of grain alignment theories. One can use the dependence of polarization fraction on column density to study grain alignment in dense cores. Previous polarimetric studies of starless cores \citep{2014A&A...569L...1A,2015AJ....149...31J} and protostars \citep{2018ApJ...855...92C} have consistently observed a decrease in polarization fraction with increasing column density, a finding referred to as “polarization hole”  \citep{2019FrASS...6...15P,2020A&A...641A..12P}. This depolarization effect has been attributed to several possible physical mechanisms, including reduced anisotropic radiation field strengths at greater depths in dense regions \citep{1992ApJ...399..108G,1995ApJ...448..748G,2016A&A...593A..87R,2016A&A...588A.129B}, increased magnetic field tangling \citep{2014ApJS..213...13H,2019MNRAS.482.2697S}, and decreased grain alignment efficiency \citep{2016A&A...588A.129B,2019ApJ...877...88C,2020A&A...644A..11L,2025A&A...703A.192T}. 

However, for the case of proto-stellar cores, RAT predicts an increase of induced polarization fraction towards the protostar as incident radiation from the protostar can enhance grain alignment efficiency \citep{Hoang_Tram_Lee_Diep_Ngoc_2021,Hoang_Tram_Minh}. Previous studies of ALMA (Atacama Large Millimeter/submillimeter Array) dust polarization data have reported enhanced grain alignment along protostellar outflow cavity walls, where irradiation is locally increased due to radiation from the central protostar and its accretion energy escaping through the cavities \citep{2018MNRAS.477.2760M,2019ApJ...885..106L,2023A&A...675A.133L}. The prediction was also observed in \cite{2020A&A...644A..11L}, which reported a constant grain alignment efficiency even at the highest intensities for high luminosity ALMA cores. While such behaviour has been observed on the scales of individual proto-stellar cores, direct observational evidence for enhanced grain alignment in star-forming regions containing multiple protostellar sources remains scarce. The prediction was partially observed in the star forming region Serpens South in \cite{2020NatAs...4.1195P}, which reported a shallow slope in the polarization fraction–intensity relation, suggesting a weaker loss of polarization in some proto-stellar cores. Even though an increasing trend in the polarization fraction-intensity was not observed, their analytical modeling predicted such a trend for the highest intensities. One of the proposed explanations for the dearth of observational evidence of enhanced grain alignment near protostars is Radiative Torque Disruption (RATD) \citep{2019NatAs...3..766H,2019ApJ...876...13H,2020ApJ...896..144H}, which proposes that high radiation intensities can cause larger grains to spin up past their tensile strengths and thus fragment into smaller grains. The depletion of large grains and the corresponding increase in the small-grain population leads to an overall reduction in net polarization fraction \citep{2020ApJ...895...16H,2020ApJ...896...44L}.

Our study aimed to test the prediction of enhanced grain alignment efficiency toward protostars from RAT alignment theory. For this purpose, DR21 was selected as the target region because it is a dense and massive filamentary structure within the Cygnus-X complex \citep{2007A&A...476.1243M,2008hsf1.book...36R,2011ApJ...727..114R} and hosts massive protostars \citep{2010A&A...524A..18B, Beerer2010, Cheng2022}, including an embedded H II region DR21 \citep{1967ApJ...148L..17R,1989A&A...222..247R} and a bright maser source DR21(OH). It is located at a distance of 1.5 kpc \citep{2012A&A...539A..79R} toward the Cygnus North region. Previously, the DR21 filament was surveyed using single-dish observations of polarized dust emission from 100 \um{} to 1.3 $\mathrm{mm}$, yielding maps of the large‐scale magnetic field morphology across parsec scales \citep{2000ApJS..128..335D,2003ApJ...598..392L,2009ApJ...694.1056K,2010ApJS..186..406D,2013ApJ...772...69G,2017ApJ...838..121C}. More recently, \cite{2022ApJ...941..122C} performed a survey of polarized dust emission in DR21 with the  \underline{S}ubmillimetre \underline{C}ommon-\underline{U}ser \underline{B}olometer \underline{A}rray 2 (SCUBA-2, \cite{2011ASPC..449...68B,2016SPIE.9914E..03F,2013MNRAS.430.2513H}. The data were obtained at  850~\um{} as part of the \underline{B}-fields \underline{I}n \underline{ST}ar-forming \underline{R}egion \underline{O}bservations (BISTRO) survey \citep{2017ApJ...842...66W} to examine the polarization fraction in DR21.

For our analysis, we used observations from the High-resolution Airborne Wideband Camera-plus (HAWC+; \citep{2007SPIE.6678E..0DV,2010SPIE.7735E..6HD,2018JAI.....740008H}) onboard the 2.7-m Stratospheric Observatory For Infrared Astronomy (SOFIA; \cite{2018JAI.....740011T}). We present 214 \um{} polarimetric maps of the DR21 region obtained with \sofia{}/\hawc{} as part of SIMPLIFI (\underline{S}tudy of \underline{I}nterstellar \underline{M}agnetic \underline{P}olarization: a \underline{L}egacy \underline{I}nvestigation of \underline{FI}laments), a legacy survey described in detail by Pillai et al. (in preparation; hereafter Paper I). In Section \ref{sec:obs}, the \sofia{}/\hawc{} observations and the data analysis pipeline are described. In Section \ref{sec:method}, the methods and tools used in this study are presented. The results are described in Section \ref{sec:results} with the inferences from the results described in Section \ref{sec:discussion}. Finally, in Section \ref{sec:conclusions}, the conclusions are collected.

\section{Observations}\label{sec:obs}

 Observations of the DR21 region were conducted using \sofia{}/\hawc{} at 214 \um{} to obtain measurements of total intensity, $I$ and linear polarization Stokes parameters, $Q$, $U$ used to parameterize the linear component of the polarization of the thermal dust emission. A full description of the SIMPLIFI survey, data processing and data products is provided in Paper I. We briefly present the data products here. In this section the focus was on essential aspects of how data from the \hawc{} instrument on \sofia{} were processed.

The \hawc{} observations were obtained as part of SOFIA project 09-0215 (PI: T.~Pillai). Observations were conducted using on-the-fly (OTF) mode during flights F776, F779, F780, F782, F784, F917, and F920 in September 2021 and September 2022. The observations were obtained in band E, a 44 \um{} wideband centered at 214 \um{}. The OTF scan speed was $200\arcsec\,{\rm{}s}^{-1}$, and the scan amplitude varied between $60\arcsec$ and $200\arcsec$. The data were acquired as a mosaic of observations centered on three central positions and were processed using version 3.2.1.dev0 of the \hawc{} Data Reduction Pipeline (DRP) \citep{2015ASPC..495..355C}. Paper I describes how the pipeline data reduction was optimized to select parameters that caused only minor cleaning of the data. The final data have a resolution of $20\farcs{}3$ (full width at half-maximum; FWHM) and map pixel size of $4\farcs{}55$. The map is centered on $(\mathrm{RA}, \mathrm{DEC}~(J2000)) = (20^{\rm h}38^{\rm m}53.40^{\rm s}, 42^{\circ}22'10.57''$)  with size of $23\arcmin \times17 \arcmin$.

The Stokes parameter maps $I, Q, U$ were used to compute the polarized intensity $P$, polarization fraction $p_{obs}$ and its associated uncertainty $\sigma_p$ \citep{2016A&A...594A..19P}:

\begin{equation}
\hspace{85pt}P = \sqrt{Q^2 + U^2},
\end{equation}
\begin{equation}
\hspace{95pt}p_{\rm{obs}} = P/I,
\end{equation}

\begin{equation}\label{eq:err_p}
\begin{split}
\hspace{10pt}\sigma^2_{p_{obs}} &= \frac{1}{p^2_{obs}I^4}\times  \Biggl[Q^2\sigma_{QQ} + U^2\sigma_{UU} + \frac{\sigma_{II}}{I}\times(Q^2 + U^2)^2 \\ 
&+ 2QU\sigma_{QU}  - 2Q\frac{(Q^2 + U^2)}{I}\sigma_{IQ} -2U\frac{(Q^2 + U^2)}{I}\sigma_{IU}\Biggr]
\end{split}
\end{equation}
where $\sigma_{AB}$, $\{A,B\}$ $\in$ $\{I,Q,U\}$, are the components of the noise covariance matrix. Specifically, $\sigma_{II}$, $\sigma_{QQ}$ and $\sigma_{UU}$ are the noise estimate on Stokes parameters I, Q and U, respectively and $\sigma_{IQ}$, $\sigma_{IU}$ and $\sigma_{QU}$ are the off-diagonal terms.

The plane of the sky polarization angle (electric vector position angle) and its associated uncertainty were calculated using the Stokes Q and U parameters \citep{2016A&A...594A..19P}:
\begin{equation}
\hspace{85pt} \psi= \frac{1}{2}\arctan2{(U,Q)},
\end{equation}

\begin{equation}
\hspace{45pt}\sigma_{\psi} = \frac{1}{2}\frac{\sqrt{(Q\sigma_{UU})^2 +(U\sigma_{QQ})^2 - 2QU\sigma_{QU}}}{Q^2 + U^2},
\end{equation}
where, the two-argument function arctan2 \footnote{arctan2 is the 2-argument arctangent. It is defined as arctan2(y,x) $=\theta$, where $\theta$ is the angle measured (in radians, $-\pi <\theta \leq \pi$) between the positive 
x-axis and a line from origin to the point (x,\,y) in the Cartesian plane.} was used to account for the $\pi$-ambiguity. The angle was subsequently rotated by 90$^\circ$ to match magnetic field direction. The debiased polarization percentage ($p$) was calculated as \citep{1974MExP...12..361S}:
\begin{equation}{\label{eq:debp}}
\hspace{85pt}p = \sqrt{p_{\rm obs}^2 - \sigma_p^2},
\end{equation}

A SNR criterion of $p/\sigma_p > 3$ was placed on the data, for which the estimator for debiased polarization fraction  (Equation~\ref{eq:debp}) performs comparably to the modified asymptotic estimator (see Figure 2 in \cite{2006PASP..118.1340V} and Figure 3 in \cite{Plaszczynski_Montier_Levrier_Tristram_2014}). In addition to a SNR cut on $p$, a SNR cut of $I/\sigma_I > 100$ was also applied, which constrains $\sigma_p \leq \sqrt{2}\times(I/\sigma_I)^{-1} \leq 1.4\%$ (approximation derived from Equation \ref{eq:err_p} by assuming small off-diagonal terms and $\sigma_{QQ}=\sigma_{UU}$). In addition to the SNR cuts, a criterion of $p<50\%$ was also applied to remove any outlier pixels with unphysical values of $p$. For the analysis in Section~\ref{sec:results}, to obtain independent data points, every other $4\farcs{}55\times4\farcs{}55$ pixel in both $\mathrm{RA}$ and $\mathrm{DEC}$ directions was selected.

For $\rm{H_2}$ column density the published, $N_{\rm{H_2}}$, map from \citet{Pokhrel2020} was used. In their work, column density and temperature maps were derived for a sample of star-forming regions using \textit{Herschel} dust emission via pixel-by-pixel, optically thin modified-blackbody fits after convolving all bands to the \textit{SPIRE} (Spectral and Photometric Imaging Receiver) \citep{griffin2010herschel} 500 \um{} beam with $37\arcsec $ FWHM resolution. For Cygnus-X (including the region of interest for this study, DR21), they fitted the intensities measured in 160, 350, and 500~\um{} bands (but not using 250 \um{} due to a mosaic background mismatch), adopted a fixed emissivity index ($\beta =1.5$), and they restored the large-scale emission using Planck-based offset calibrations.

\section{Methods}\label{sec:method}
In this paper, a statistical analysis of the dust polarization fraction in DR21 was performed, examining its dependence on key physical properties including intensity, turbulent magnetic field, and grain alignment efficiency. To study the dependence of polarization fraction on intensity, potential biases arising from the loss of low-surface-brightness emission had to be accounted for. This bias was assessed by examining the flux recovery of \sofia{}/\hawc{} through comparison with a Planck-corrected {\it Herschel}/PACS 250 \um{} map \citep{Pokhrel2020}. To incorporate the effects of turbulent magnetic fields and grain alignment efficiency into the analysis, these quantities were quantified using observations of polarization fraction and polarization angle. The level of magnetic field turbulence was characterized using the angular dispersion of polarization angles \citep{2015A&A...576A.104P}. The product of polarization fraction and angular dispersion, $p \times S$ \citep{2020A&A...641A..12P}, was used as a proxy for the intrinsic dust polarization efficiency, which partially accounts for depolarization due to magnetic field tangling along the line of sight. These methodological steps are described in more detail in the following sections.

\subsection{Flux recovery by \sofia{}/\hawc{}}
\begin{figure*}[h!]
    \centering
    \includegraphics[width=0.9\linewidth]{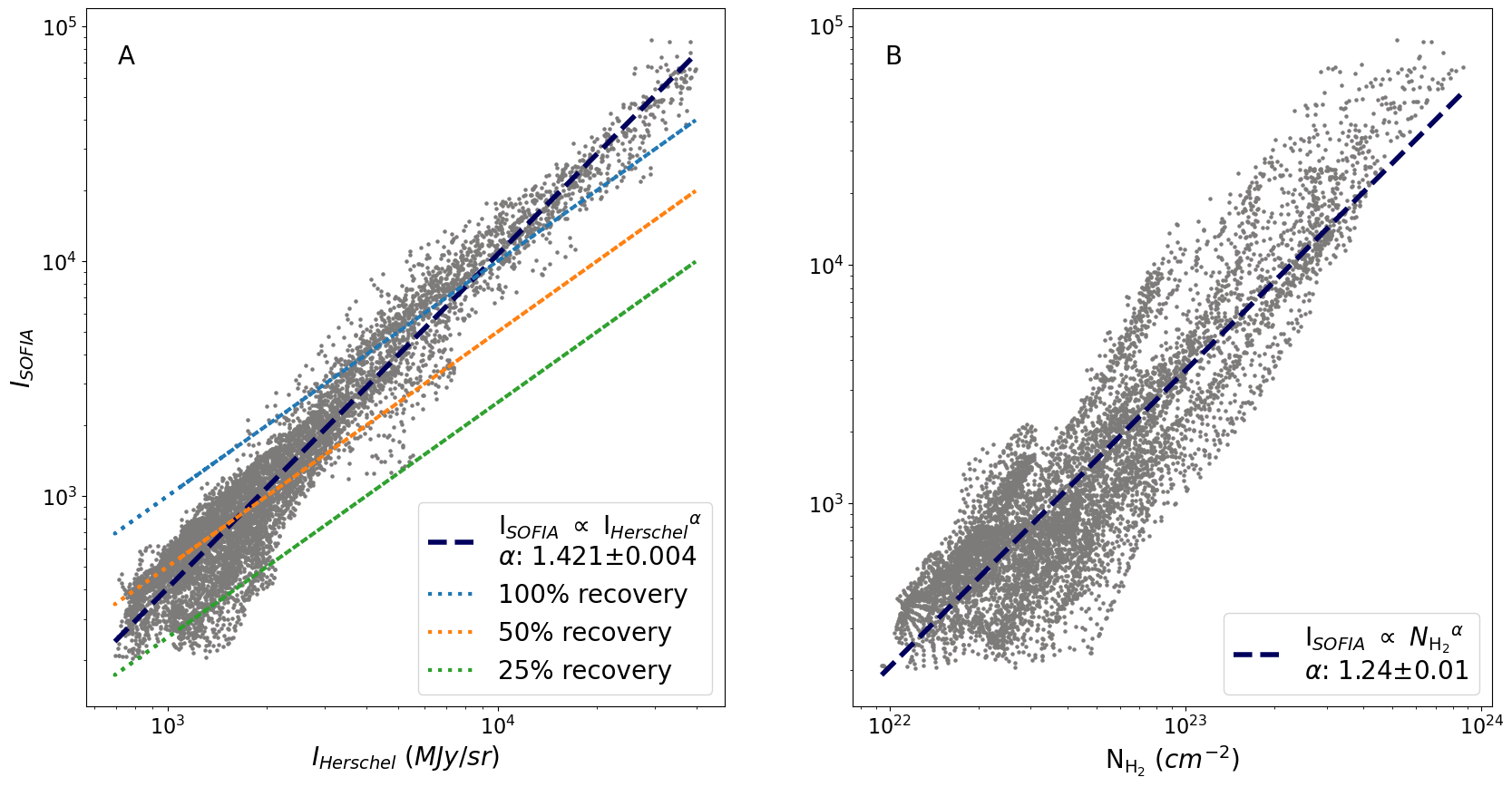}
    {\caption{Comparison of observed \sofia{} 214\,\textmu m brightness \(I_{\sofia{}}\) versus  250\,\textmu m \textit{Herschel} brightness \(I_{Herschel}\) (left A panel) and $N_{\rm H_2}$ (right panel B). In panel A, blue, orange and green dotted lines represent the reference line for 100\,\% ($I_{\sofia{}}$  = $I_{ Herschel}$), 50\,\% ($I_{\sofia{}}$ = $0.5I_{ Herschel}$) and 25\,\% ($I_{\sofia{}}$ = $0.25 I_{Herschel}$) recovery, respectively.  The dark blue dashed line is the best‐fit \(I_{\sofia{}}\propto I_{ Herschel}^{1.421\pm 0.004}\). Similarly in panel B, the dark-blue dashed line is the best‐fit \(I_{\sofia{}}\propto I_{N_{\rm H_2}}^{1.24\pm 0.01
    }\). \label{fig:flux-recovery}}}
\end{figure*}

\label{sec:flux-recovery}

Obtained at stratospheric altitudes, the \sofia{}/\hawc{} data required correction for instrumental offsets, telescope emissivity and residual atmospheric “sky” emission. This was achieved either through chop–nod or On-The-Fly (OTF) mapping (the latter was the case for our DR21 observations) that included periodic blank‐sky measurements (see Paper I). However, these observing modes inevitably filter out astrophysical emission on angular scales comparable to the chop throw or the OTF scan length. Since the polarization fraction, $p$, is derived from both the total intensity ($I$) and polarized intensity ($P$), loss of sensitivity to low–surface‐brightness structure could bias $p$ measurements. The Planck‐corrected {\it Herschel} 250\,\um{} map \citep{Pokhrel2020}, regridded to the \sofia{} beam and pixel scale (20\farcs{}3 and 4\farcs{}55) was used as the starting point and assumed to be a good proxy for emission at 214 \um{}. 

Figure~\ref{fig:flux-recovery} compares the OTF-recovered SOFIA 214~\um{} intensity with \textit{Herschel} 250 \um{} intensity (panel A) and with $\mathrm{H_2}$ column density (panel B). Panel A shows the \sofia{} intensity, \(\IS\) versus \textit{Herschel} intensity, \(\IH\) with reference lines for 100\% recovery (\(\IS=\IH\)), 50\% recovery (\(\IS=0.5\,\IH\))  and 25\% recovery (\(\IS=0.25\,\IH\)) overploted. A least‐squares fit to \(\IS \propto \IH^\alpha\) yields \(\alpha = 1.421\pm0.004\), as shown by the dark-blue dashed line. The superlinear index (\(\alpha>1\)) indicates that low‐intensity, extended emission was disproportionately attenuated by the SOFIA OTF sky subtraction, whereas bright emission from compact regions was less affected. Quantitatively, \sofia{} recovered \(\ge50\%\) of the \textit{Herschel} emission for \(\approx63\%\) of the mapped pixels, and between 25\%–50\% for \(\approx33\%\) of the pixels, implying that faint, extended structure was suppressed in the \sofia{} maps. At the highest intensities, a small cluster of points lies above the one‐to‐one line, coincident with high‐mass YSOs; there, the assumption of identical 214\,\um{} and 250\,\um{} beam behaviour breaks down, and partial saturation in the {\it Herschel}/SPIRE detectors likely underestimates \(\IS\).  Similar to our results, \citet{Coude2025} found that the discrepancies between \textit{Herschel} and \sofia{}/\hawc{} fluxes cannot be explained by a simple constant offset, but rather reflect both background subtraction effects at low intensities and beam dilution or saturation effects for \textit{Herschel} at the high-flux end.

In panel B of Figure \ref{fig:flux-recovery}, the \sofia{} 214\,\um{} intensity is plotted versus the {\it Herschel} based column density \(N_{\rm H_2}\) from \cite{Pokhrel2020}. A power‐law fit \(\  \IH \;\propto\; N_{\rm H_2}^{\,1.2\pm0.01}\) shows that the bulk of the \hawc{} emission traces the total H\(_2\) column.

\subsection{Angular dispersion and a proxy for dust grain alignment efficiency\label{subsec:ang_dis}}

\cite{2015A&A...576A.104P} quantified the depolarization effect caused by tangled magnetic fields using a statistical analysis of polarization fractions and dispersions of plane of the sky polarization angles. In this Section, the method described in Section 3.3 of \cite{2015A&A...576A.104P} was followed. The angular dispersion function is defined as:
\begin{equation}\label{eq:adf}
\hspace{75pt}S_b(x,\delta) = \left( \frac{1}{N}\sum_{i = 1}^{N}(\Delta \psi_{xi})^2  \right)^{1/2}
\end{equation}
where $\Delta\psi_{xi} = \psi(x) - \psi(x + \delta_i)$ is the angle difference between $\psi(x)$ and $\psi(x + \delta_i)$, the polarization angle at a given position of the sky $x$ (central pixel) and the polarization angle at a pixel position of the sky displaced from the position $x$ by the displacement vector $\delta_i$, respectively. The average in Equation~ (\ref{eq:adf}) spans $N$ pixels indexed by $i$ and located at positions $x + \delta_i$ within an annulus centered on $x$, with inner and outer radii of $\delta/2 $ and $3\delta/2 $, respectively. We set $\delta$ to the beam size 20\farcs{}3. The chosen $\delta$ allows for at most 5 \sofia{} beams per annulus.

The angular difference, $\Delta\psi_{xi}$ was computed using the full Stokes $Q$ and $U$ maps as:

\begin{equation}\label{eq:del_psi}
\hspace{32pt} \Delta\psi_{xi} = \frac{1}{2}\mathrm{arctan2}(Q_i U_x - Q_x U_i, Q_i Q_x + U_i U_x),
\end{equation}
where the indices $x$ and $i$ represent the central pixel and the pixel in the displaced position, respectively, and where the two-argument function arctan2 was used to take care of the $\pi$-ambiguity.

As $S_b$ is positively biased in the presence of noise, it needs to be debiased by removing the bias, as described in \cite{2009ApJ...696..567H}:

\begin{equation}\label{eq:deb_s}
\hspace{75pt}
S^2(\delta) = S_b^2(\delta) - \sigma_{S_b}^2(\delta) 
\end{equation}
The above equation is a good approximation for debiased $S$ in the case of $S_b/\sigma_S > 3$ \citep{2020A&A...641A..12P}, and accordingly, this SNR cut on $S$ was applied. The noise was calculated as described in \cite{2020A&A...641A..12P}:
\begin{equation}\label{eq:sigma_s}
\begin{split}
 \hspace{40pt} \sigma^{2}_{S_b}(\delta) & = \frac{\sigma^2_{\psi}(x)}{N^2 S_b^2}\left(\sum_{i=1}^{N}\Delta\psi_{xi}\right)^2 \\
 & + \frac{1}{N^2 S_b^2}\sum_{i=1}^{N}\sigma_{\psi}^2(x + \delta_i)(\Delta\psi_{xi})^2,   
\end{split}
\end{equation}
where $\sigma_{\psi}(x)$ and $\sigma_{\psi}(x + \delta_i)$ are the error in polarization angle at position $x$ and $i$ respectively and $N$ is the same as defined in Equation~(\ref{eq:adf}). The debiased angular dispersion provides a measure of the disorder in the magnetic field projected onto the plane of the sky. In regions of uniform polarization position angle, $S$ is low, while in regions where the polarization position angle is more randomly distributed, $S$ is higher. For a completely random distribution of an infinite number of polarization angles, $S$ saturates at a theoretical upper limit $\pi/\sqrt{12}\approx52^{\circ}$ \citep{2016A&A...595A..57A} with any measurement above this value being unreliable. Therefore, to use angular dispersion for analysis, the saturation limit for the data must first be determined.

\begin{figure*}[htb!]
\centering
\includegraphics[width=0.8\linewidth]{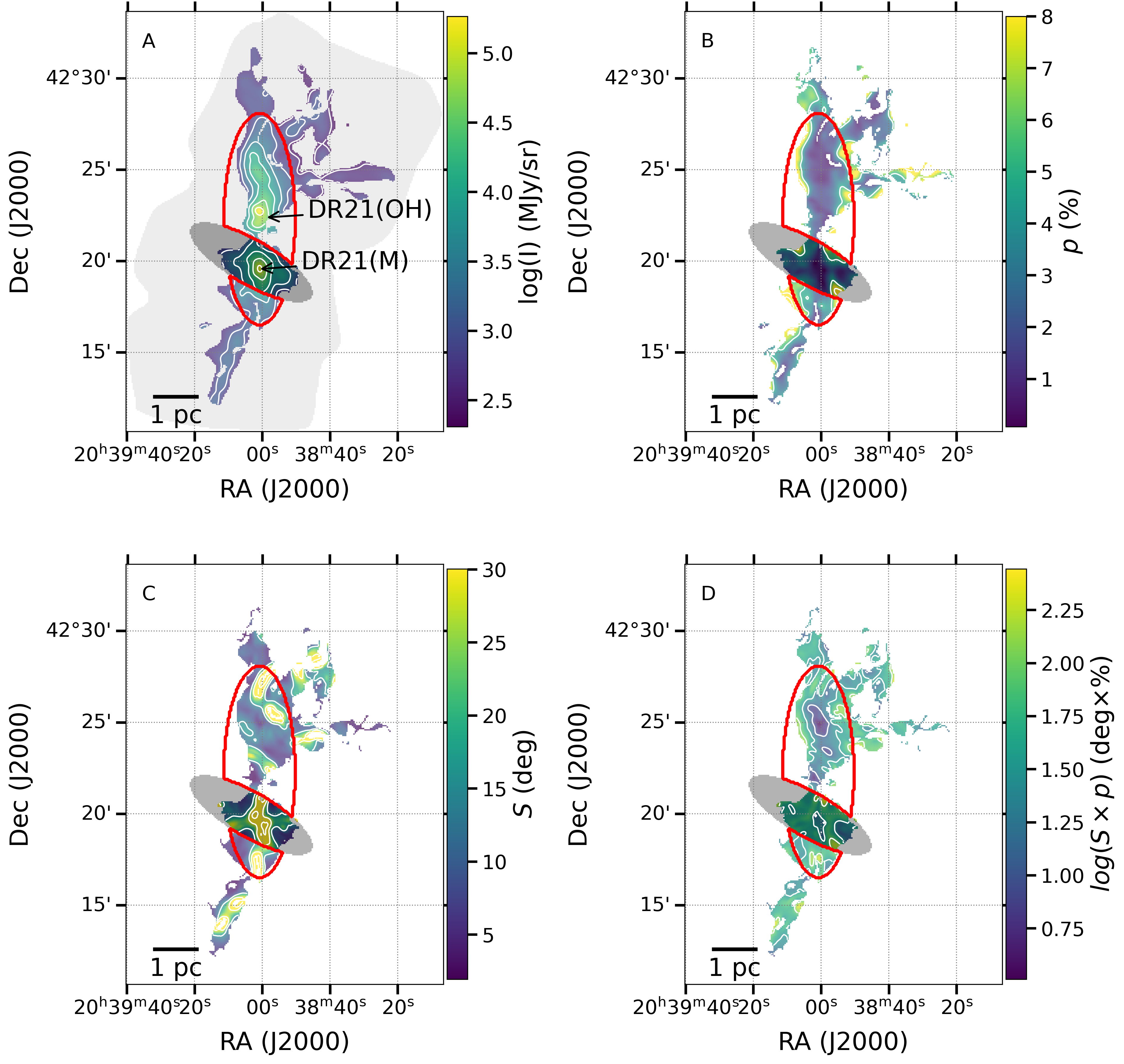}
\caption{Main data products derived from \sofia{}/\hawc{} observations used in the study, after application of quality cuts. The \sofia{}/\hawc{} map footprint is shown as a light shadow in panel A. Panel A: map of Stokes $I$, panel B: map of polarization fraction $p$, panel C: map of debiased angular dispersion $S$, with $\delta = 20\farcs{}3$ and panel D: map of $S\times p$. The two massive proto-stellar regions, DR21(OH) and DR21(M) are labeled in panel A. The DR21 Ridge is inside the broken ellipse outlined in red, the DR21 sub-filaments are located outside of the red ellipse. The mask for the outflow is shown as a gray, filled ellipse.
\label{fig:main_data}}
\end{figure*}

The saturation limit for angular dispersion was estimated as follows. Consider a distribution of angles $k_i$ drawn from a Gaussian distribution with mean zero and standard deviation $\sigma_{k}$. For the analysis, the "unwrapped" angles were folded back into the range $-90^\circ$ to $+90^\circ$, which is equivalent to the angles $\phi_i=\arcsin(\sin[k_i])$ considereding Monte Carlo experiments with large samples of $k_i$  showed that the standard deviation of $\phi_i$, denoted $\mathbb{S}$ (analogous to $S$ used in this study), remains within 10\% of $\sigma_{k}$ when $\mathbb{S}<45^\circ$. The choice of 10\% is not critical since the relationship between $\mathbb{S}$ and $\sigma_{k}$ exhibits a sharp knee near saturation, such that varying the threshold from 5\% to 20\% shifts the cutoff only from $\sim40^\circ$ to $\sim49^\circ$, a narrow range compared to the uniform-distribution ceiling. Any reasonable threshold yields essentially the same practical limit on $\mathbb{S}$. 

In other words, for a sufficiently large sample of $\phi_i$ values, measurements of $\mathbb{S}$ are reliable and unsaturated when $\mathbb{S}<45^\circ$.
However, the samples $\phi_i$ may be of limited number. This means that $\mathbb{S}$ derived from $N$ observations of $\phi_i$ may differ from the true value that would be obtained for $N\to\infty$. The annular ring
in our dispersion angle calculation corresponds to $N \leq 5.56$ independent SOFIA beams. Thus for the case $N=5$, an ensemble of 10,000 trial values of $\mathbb{S}$ were generated and their spread quantified using the standard deviation of $\mathbb{S}$, $\sigma(\mathbb{S})$. To ensure that measurements of $S$ remain unsaturated, we therefore require $S<45^\circ-\sigma(\mathbb{S}) = 29.2^{\circ}$.

\cite{2020A&A...641A..12P} argued that angular dispersion ($S$) combined with polarization fraction ($p$) can also be used as a proxy for grain alignment efficiency. They assumed that the total emission arises from a small number, $N$, of independent layers, each contributing a fraction $1/N$ of the total intensity. The magnetic field was modeled as the sum of a uniform component and an isotropic turbulent component. Under these assumptions, they derived an analytical expression relating the angular dispersion and the polarization fraction:
\begin{equation}\label{eq:adf_analytical}
   \hspace{75pt} \left<S(\delta)\right>_p \approx \frac{f_m(\delta)}{\sqrt{N}}\frac{p_{max}}{p},
\end{equation}
where $\left<S(\delta)\right>_p$ is the average angular dispersion of the map pixels with polarization fraction $p$, $N$ is the number of layers in their model, $f_m(\delta)$ depends on the properties of the turbulence on
 the scale of $\delta$ ($\delta$ is the same as defined in Equation~(\ref{eq:adf})) and $p_{max}$ is the maximum polarization that the dust grains can generate. The analytical result implies that $S\propto p^{-1}$. When magnetic field disorder increases, $S$ increases and in turn $p$ decreases, and vice-versa. Using Equation~(\ref{eq:adf_analytical}), the product $S\times p$ removes the impact of the magnetic field structure statistically and can be utilized as an empirical proxy for grain alignment efficiency \citep{2020A&A...641A..12P}. However, it must be noted that it is not a direct measure of grain alignment efficiency as the product also depends on other dust properties as well.

\subsection{Statistical tools}\label{sec:method_stat}
To explore trends and relationships between different physical properties, power-law models were fitted in log–log space:
\begin{equation}\label{eq:singlefit}
\hspace{75pt} \log\hspace{2pt} Y = C + \alpha \log\hspace{2pt} X,
\end{equation}
where $X$ and $Y$ represent the physical properties under study ($\{X,Y\}$ $\in$ $\{p,I,S,N_{\rm H_2},S\times p\}$).
The LMFIT Python package \citep{newville_2014_11813} was used to fit the data. In addition to a single power-law model, broken power-law models with one, two, and three breaks, where the break positions were treated as free parameters, were also fitted.
The best model to represent each trend was selected based on its Bayesian Information Criterion (BIC) value \citep{1978AnSta...6..461S}.

\begin{figure*}[htb!]
\centering
\includegraphics[width=0.8\linewidth]{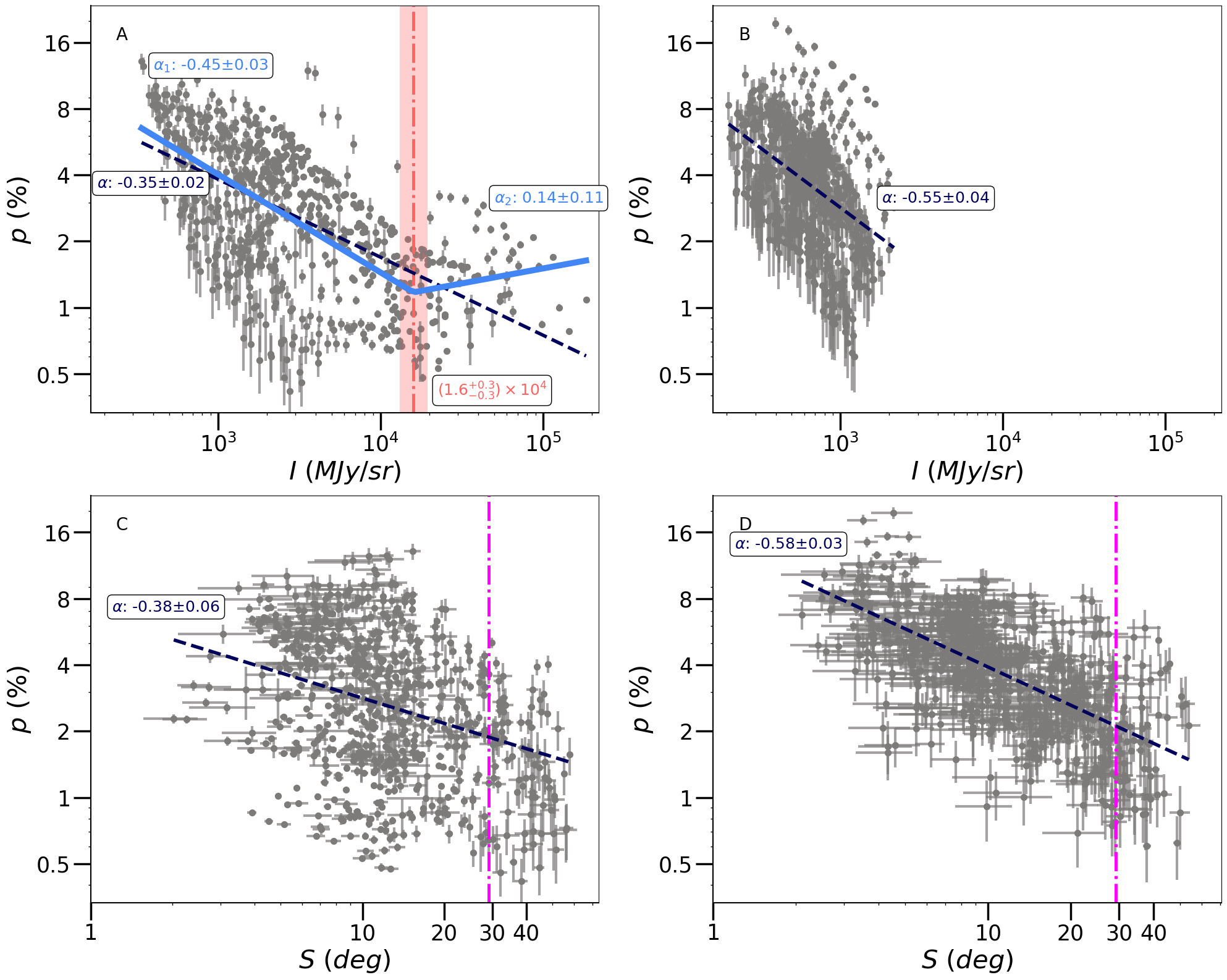}
\caption{Trends of $p$ versus $I$ (top) and $p$ versus $S$ (bottom). The data points are plotted in gray with the Ridge in panels A and C and the sub-filament pixel values plotted in B and D. The single power-law fit is shown as a dashed dark blue line in all four panels. In the case of $p$ vs $I$, for the Ridge (panel A), a broken power-law fit (solid light blue) is also included. The break in intensity  is marked by a vertical red dot-dashed line and the error in the break point location is shown as a red shading ($\pm1\sigma$) around the vertical line. In the case of $p$ vs $S$, for both the plots (bottom), the saturation limit of angular dispersion is marked by a magenta dot-dashed line at $29.2^{\mathrm{o}}$.\label{fig:p_vs_S_and_I}}
\end{figure*}

\begin{figure*}[htb!]
\centering
\includegraphics[width=0.8\linewidth]{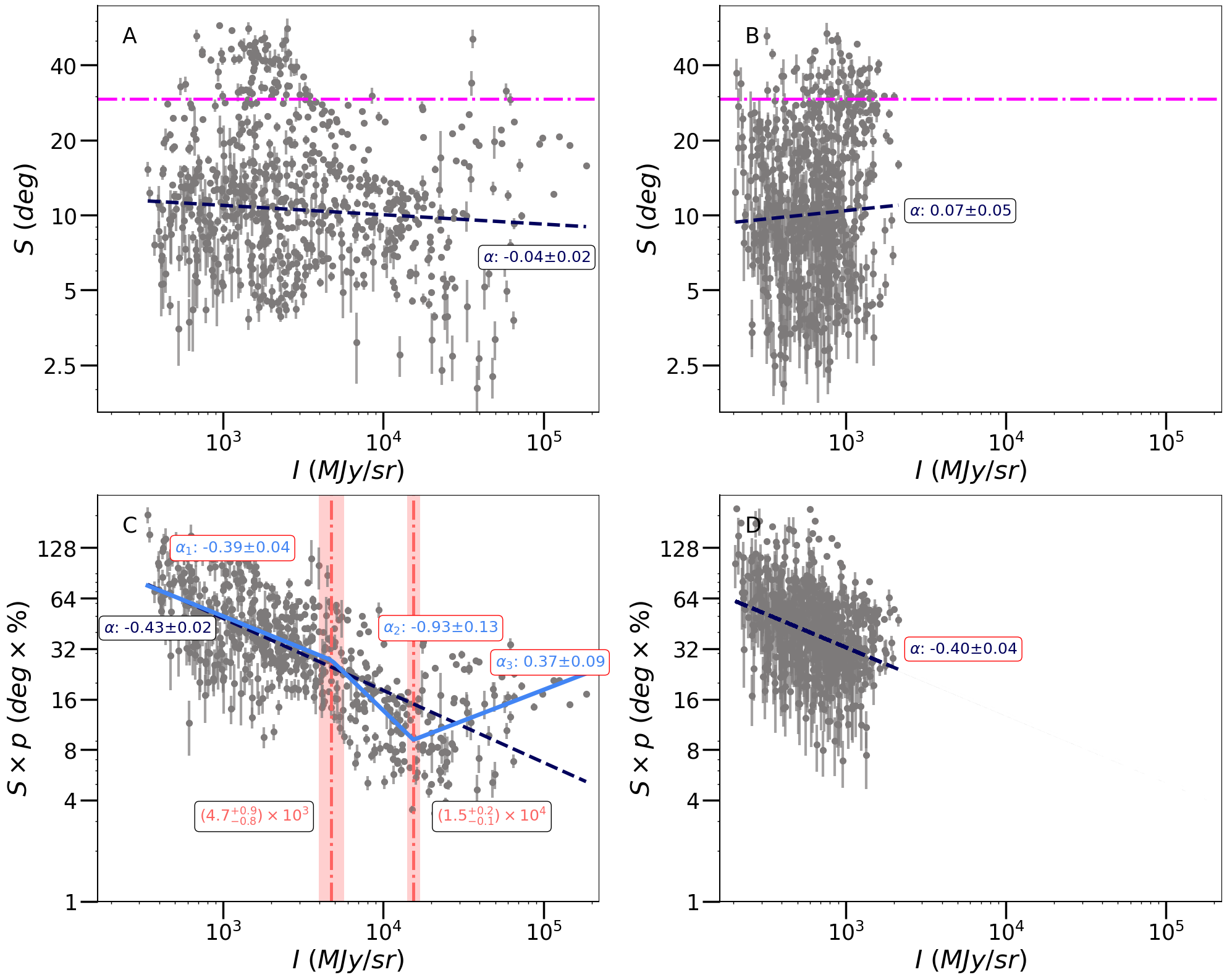}
\caption{Similar to Figure \ref{fig:p_vs_S_and_I} but showing the relationship between $S$ versus $I$ (panels A and B) and $S\times p$ versus $I$ (panels C and D). The data points are plotted in gray with the Ridge in panels A and C and the sub-filaments in B and D. The saturation limit of the angular dispersion is marked by a magenta dot-dashed line at $29.2^\circ$. In the $S\times p$ plot for the Ridge (panel C), in addition to a single power law, a two break broken power law is fitted shown in solid light blue. The breaks are marked by vertical red dot-dashed lines, with error ranges ($\pm1\sigma$) shown with red shading \label{fig:s_vs_I_sp_vs_I}}.
\end{figure*}

\section{Results}\label{sec:results}
Figure~\ref{fig:main_data} presents the main data products extracted from the SOFIA/HAWC+ observations  used in this study of DR21: the map of Stokes $I$ (panel A), polarization fraction $p$ (panel B), angular dispersion $S$ (panel C), and $S\times p$ (panel D). In the Stokes $I$ map (panel A), two prominent intensity peaks are visible, corresponding to the maser DR21(OH) in the north and the embedded HII region DR21(M) in the south \citep{1967ApJ...148L..17R,1989A&A...222..247R,2010A&A...524A..18B,2010A&A...520A..49S}. \cite{Ching_Qiu_Li_Ren_Lai_Berry_Pattle_Furuya_Ward-Thompson_Johnstone_et} described the star-forming region DR21 as made up of a main filament (referred to hereafter as the Ridge) and several smaller filamentary structures (the F1, F3, SW, and S sub-filaments) surrounding the main filament.

To better understand the relevant physics in these sub-structures, we divided the DR21 region into two main sub-regions similar to \cite{Ching_Qiu_Li_Ren_Lai_Berry_Pattle_Furuya_Ward-Thompson_Johnstone_et}: (1) The Ridge (outlined by the red contour in Figure \ref{fig:main_data}), and (2) the surrounding sub-filaments located outside this contour. In addition to these two sub-regions, there also exists a strong molecular outflow around the embedded HII region DR21(M) \citep{1996A&A...310..961D,2010A&A...518L.114W}. This outflow is shown as a shaded area (gray filled ellipse) in all panels. The motivation behind the regions' definition is described in Paper I. The outflow, characterized as highly energetic in that work, was masked out in the present RAT-focused study as energetic outflows might trigger non-RAT alignment mechanisms which were not the subject of this study \citep[see][for a review]{2015ARA&A..53..501A}.

A qualitative inspection of the maps in Figure~\ref{fig:main_data} reveals several key trends. Regions with high Stokes $I$ generally exhibit lower polarization fractions, consistent with the well-known depolarization effect \citep{2016ApJ...824..134F,2020NatAs...4.1195P,2020A&A...644A..11L}. Most areas of high angular dispersion coincide with regions of low Stokes $I$, although a patch of moderate levels of $S$ is seen near DR21(OH). The $S\times p$ map similarly suggests high grain alignment efficiency in low-intensity regions with the exception of moderate levels of $S\times p$ near DR21(OH), likely reflecting local effects from embedded sources.

In this study, two explanations for the observed depolarization trend were explored: (1) tangling of the magnetic field \citep{2014ApJS..213...13H,2019MNRAS.482.2697S}, and (2) a decrease in dust grain alignment efficiency \citep{2016A&A...588A.129B,2019ApJ...877...88C,2025A&A...703A.192T}. To test these, the relationships among polarization fraction, intensity, and angular dispersion as tracers of magnetic field disorder were examined. Furthermore, dust grain alignment efficiency as a function of intensity was evaluated.

Figure \ref{fig:p_vs_S_and_I} shows polarization fraction versus intensity in the top panels and the angular dispersion versus intensity in the bottom panels. Panels A and B of Figure \ref{fig:p_vs_S_and_I} compare the values of polarization fraction ($p$) and intensity ($I$) for the DR21 Ridge (panel A) and the sub-filaments (panel B). Both regions exhibit anti-correlations between $p$ and $I$. Power-law relations were fitted to the data in each region, yielding $p \propto I^{-0.35\pm0.02}$ (Spearman correlation coefficient ($\rho$) = $-0.71$ and p-value $\ll 0.001$) for the Ridge and $p \propto I^{-0.55\pm0.04}$ ($\rho = -0.39$, p-value $\ll 0.001$) for the sub-filaments. The Ridge shows a shallower power law index as compared to the sub-filaments, suggesting a weaker depolarization effect. A BIC comparison of the broken power-law models showed that a broken power-law model with one break provided the best fit for the Ridge, while the sub-filaments were adequately described by a single power-law. For the Ridge, the broken power-law fit identifies a break at an intensity of $1.6^{+0.3}_{-0.3} \times 10^4$ MJy/sr. Below this, the index $\alpha = -0.45\pm 0.03$ is shallower than that for the sub-filaments, while above this threshold the slope almost flattens ($\alpha = 0.14\pm 0.11$), indicating an approximately constant polarization fraction. It must be noted that the estimation of $\alpha$ can be biased by data points in the low S/N regime. This bias was explored in Appendix \ref{sec:bias_in_alpha} and shows that it is likely negligible for our dataset.

Panels C and D of Figure \ref{fig:p_vs_S_and_I} compare the values of polarization fraction ($p$) and angular dispersion ($S$) for the DR21 Ridge (panel C) and the sub-filaments (panel D). Data points above the angular dispersion saturation limit of $29.2^\circ$ (magenta line in panel C of Figure \ref{fig:p_vs_S_and_I}) were not included in the analysis. A negative correlation was observed between polarization fractions and angular dispersions. Power-laws (Equation~(\ref{eq:singlefit})) were fitted to the data in each region, yielding relatively shallow
slopes of $p \propto S^{-0.38\pm0.06}$ ($\rho = -0.28$, p-value $\ll 0.001$) for the Ridge and $p \propto S^{-0.58\pm0.03}$ ($\rho = -0.62$, p-value $\ll 0.001$) for the sub-filaments. A single power-law model was found to be the best fit for both the Ridge and sub-filaments based on the BIC. 
The polarization fraction at the lowest angular dispersion is higher in the sub-filament than in the Ridge. However the power law fit indicates a steeper slope for the sub-filaments. This suggests that the effects of field tangling on depolarization are more pronounced in the sub-filaments than in the Ridge. These trends are further discussed in Section \ref{sec:discussion}.

Figure \ref{fig:s_vs_I_sp_vs_I} shows angular dispersion versus intensity in the top panels and $S\times p$ (a proxy for grain alignment efficiency) versus intensity in the bottom panels. The left column shows the pixel values for the Ridge, while the right column shows the same for the sub-filaments. Each relation was fitted with a power law. A broken power law was also used for the $S\times p$ versus $I$ relation in the Ridge.

Panels A and B of Figure~\ref{fig:s_vs_I_sp_vs_I} compare the values of angular dispersion ($S$) and intensity ($I$), with the Ridge shown on the left and the sub-filaments on the right. For this analysis, only angular dispersion values below the saturation limit were included in the fitting. High-density regions within DR21 correspond to star-forming structures where local processes such as the enhanced influence of self-gravity or turbulence driven by localized feedback, are generally expected to introduce greater disorder in the magnetic field orientations, and thus yield larger values of $S$. However, on scales of $\sim0.2$ pc (for the distance of 1.5 kpc and angular resolution of 20\farcs{}3), our analysis reveals no significant increase in the dispersion $S$ with rising $I$. Instead, the angular dispersion displays a broad distribution across all intensity ranges. A power-law fit to the data using Equation~(\ref{eq:singlefit}) yields $S\propto I^{-0.04\pm0.02}$ ($\rho = -0.09$, p-value$=0.03$) for the Ridge and $S\propto I^{0.07\pm0.05}$ ($\rho = 0.04$, p-value$=0.24$) for the sub-filaments, indicating a negligible correlation. Hence angular dispersion was found to be uncorrelated with intensity when modeled as a single power law with the single power-law model being the best model based on BIC for both the Ridge and the sub-filament.

Panels C and D of Figure \ref{fig:s_vs_I_sp_vs_I} compare the values of $S \times p$ and intensity ($I$) for the Ridge (panel C) and the sub-filaments (panel D). A negative correlation was found between $S\times p$ and intensity. The single power-law fit yielded $S\times p \propto I^{-0.43\pm0.02}$ ($\rho =-0.74$, p-value $\ll 0.001$) for the Ridge and $S\times p \propto I^{-0.40\pm0.04}$ ($\rho =-0.36$, p-value $\ll 0.001$)  for the sub-filaments. Since the angular dispersion $S$ was shown above to be uncorrelated with $I$, variations in $S \times p$ mainly reflect changes in polarization fraction $p$ with intensity (panels A and B of Figure \ref{fig:p_vs_S_and_I}).

In addition to a single power law,  broken power-law models with one, two, and three breaks were each fitted for the Ridge. Based on the BIC comparisons, a two-break model was selected as the best. The break points were estimated to be at $I = 4.7^{+0.9}_{-0.8} \times 10^3$ and $1.5^{+0.2}_{-0.1} \times 10^4$\,MJy/sr. For data points below the first break point ($I < 4.7 \times 10^3$ MJy/sr), the slope was found to be $\alpha = -0.39\pm0.04$. In the intermediate intensity regime, between $5.3 \times 10^3$ and $1.5 \times 10^4$ MJy/sr, the negative correlation steepens, yielding a slope of $\alpha = -0.93\pm 0.13$. At intensities above $1.5 \times 10^4$ MJy/sr, the trend reverses, and a positive slope of $\alpha = 0.37\pm0.09$ emerges. This turn-up of the grain alignment proxy at high intensities indicates an increasing dust-grain alignment efficiency. The analysis was repeated using $\mathrm{H_2}$ column density instead of intensity and yielded a consistent result namely a positive slope at the highest column density (Appendix \ref{sec:app_vsN}). The results on the slopes and break points of the various relations are summarized in Table \ref{tab:power_law_fits}.

\begin{table*}
\centering
\caption{Power law fitting results for polarization trends.\newline R = Ridge; SF = Sub-filaments; \ldots = Not applicable. 
$N_{\mathrm{break}}$ is the number of breaks in the broken power law fit. 
Break values are in MJy sr$^{-1}$. 
The $\alpha_1$, $\alpha_2$, and $\alpha_3$ refer to slopes before, between, and after the break(s), respectively.}
\label{tab:power_law_fits}
\setlength{\tabcolsep}{4pt}
\renewcommand{\arraystretch}{1.4}
\resizebox{\textwidth}{!}{%
\scriptsize
\begin{tabular}{l c c ccc cc}
\hline\hline
 & \multicolumn{1}{c}{Single Power Law} & \multicolumn{6}{c}{Broken Power Law} \\
\cmidrule(lr){2-2} \cmidrule(lr){3-8}
Trend & $\alpha$ & $N_\mathrm{break}$ & Break$_1$ & Break$_2$ & $\alpha_1$ & $\alpha_2$ & $\alpha_3$ \\
      &          &                    & (MJy/sr)  & (MJy/sr)  &            &            &            \\
\hline
R: $p$--$I$          & $-0.35\pm0.02$ & 1 & $1.6^{+0.3}_{-0.3}\times10^4$ & \ldots & $-0.45\pm0.03$ & $0.14\pm0.11$ & \ldots \\
SF: $p$--$I$         & $-0.55\pm0.04$ & 0& \ldots & \ldots & \ldots & \ldots & \ldots \\
R: $p$--$S$          & $-0.38\pm0.06$ & 0& \ldots & \ldots & \ldots & \ldots & \ldots \\
SF: $p$--$S$         & $-0.58\pm0.03$ & 0 & \ldots & \ldots & \ldots & \ldots & \ldots \\
R: $S$--$I$          & $-0.04\pm0.02$ & 0 & \ldots & \ldots & \ldots & \ldots & \ldots \\
SF: $S$--$I$         & $0.07\pm0.05$  & 0 & \ldots & \ldots & \ldots & \ldots & \ldots \\
R: $S\times p$--$I$  & $-0.43\pm0.02$ & 2 & $4.7^{+0.9}_{-0.8}\times10^3$ & $1.5^{+0.2}_{-0.1}\times10^4$ & $-0.39\pm0.04$ & $-0.93\pm0.13$ & $0.37\pm0.09$ \\
SF: $S\times p$--$I$ & $-0.40\pm0.04$ & 0 & \ldots & \ldots & \ldots & \ldots & \ldots \\
\hline
\end{tabular}}
\end{table*}

\section{Discussion}\label{sec:discussion}

In this section, we test predictions from RAT grain alignment theory against the observational trends identified in Section~\ref{sec:results}.

\subsection{Dust grain alignment as a function of Intensity}

In Section \ref{sec:results}, the variations in polarization fraction ($p$) as a function of intensity ($I$) across the DR21 Ridge and its associated sub-filaments were examined (see panels A and B of Figure \ref{fig:p_vs_S_and_I}).
An anti-correlation was observed in both regions upto ($> 1.6^{+0.3}_{-0.3} \times 10^4$ MJy/sr) while above that a flattening of the trend was observed for the Ridge. The power-law slope for $p$ versus $I$ at lower intensities (below $1.6^{+0.3}_{-0.3} \times 10^4$ MJy/sr) is consistent with values observed in previous studies of the polarization fraction in pre-stellar and proto-stellar cores \citep{2014A&A...569L...1A,2014ApJS..213...13H,2016ApJ...824..134F,2018ApJ...855...92C,2019FrASS...6...15P,2020A&A...641A..12P}. A previous study of DR21 \citep{Ching_Qiu_Li_Ren_Lai_Berry_Pattle_Furuya_Ward-Thompson_Johnstone_et} also reported decreasing polarization fractions as a function of intensity. This observed depolarization in dense cores has generally been explained by invoking the RAT alignment theory \citep{2007MNRAS.378..910L,2008MNRAS.388..117H}.

In this study, loss of grain alignment efficiency and tangled magnetic fields were explored in the context of RAT as possible explanations for the depolarization with increasing intensity (below $1.6^{+0.3}_{-0.3} \times 10^4$ MJy/sr), as well as for the observed flattening of the polarization fraction at the highest intensities (above $1.6^{+0.3}_{-0.3} \times 10^4$ MJy/sr).

The depolarization due to the tangling of magnetic fields was evaluated using the angular dispersion $S$ and the results were summarized in Section \ref{sec:results} (see panels C and D of Figure \ref{fig:p_vs_S_and_I}). The inverse relationships between $p \propto S^{-0.38\pm0.06}$ in the Ridge and $p \propto S^{-0.58\pm0.03}$ in the sub-filaments have similar values as observed for other clouds \citep{2016ApJ...824..134F,Ngoc_Hoang_Diep_Tram_2024}. Although an anti-correlation between $p$ and $S$ was observed, $S$ remains relatively uniform as a function of $I$ (panels A and B of Figure \ref{fig:s_vs_I_sp_vs_I}) in both the Ridge and sub-filaments. A similar result of uniform angular dispersion as a function of column density has been observed at cloud scales \citep[e.g.][see their Figure~13]{2020A&A...641A..12P}.
\cite{2020A&A...644A..11L} also reported uniform $S$ as a function of  $\rm N_{H_2}$ at core scales (see their Figure~5), with a change to increasing $S$ with column density observed only at the highest column densities. Since magnetic field disorder does not increase at higher intensity/column density in DR21, the flattening at high intensities cannot be solely due to field tangling, implying that grain alignment efficiency is a key driver of the $p$ versus $I$ trend.

In this study, the product $S\times p$ was used as a proxy for grain alignment efficiency \citep[following e.g.][]{2015A&A...576A.104P,2020A&A...644A..11L}. In Section \ref{sec:results}, the global fits to $S\times p$ versus $I$ both in the Ridge (for intensities less than $I = 1.5^{+0.2}_{-0.1}\times10^4$ MJy/sr) and the sub-filaments show negative correlations, indicating that grain alignment efficiency decreases with increasing intensity. This is consistent with the predictions from RAT alignment theory \citep{2025A&A...703A.192T}. In the Ridge, however, breaks in the $S\times p$ trend were identified, where the negative correlation first becomes steeper between $I = 4.7^{+0.9}_{-0.8}\times10^3$ and $1.5^{+0.2}_{-0.1}\times10^4$ MJy/sr and then reverses to become positive at greater intensities. This change above $1.5^{+0.2}_{-0.1}\times10^4$ MJy/sr corresponds to the flattening seen in the $p$ versus $I$ relation (above $1.6^{+0.3}_{-0.3}\times10^4$ MJy/sr), suggesting that the flattening is caused by an increase in grain alignment efficiency. This may result from a local enhancement of grain alignment efficiency due to strong anisotropic radiation from embedded proto-stars, a possibility that is further examined in the sections below.

In this study, grains are assumed to be aligned with the magnetic field (B-RAT) when rotating the polarization vectors by 90 degrees. Alternatively, the grains could be aligned with the radiation field direction (k-RAT; \citealt{2007MNRAS.378..910L,Tazaki_Lazarian_Nomura_2017,Pattle2021}) and explain the observed increase in grain alignment efficiency. However, the multi-wavelength \sofia{}/\hawc{} study of the Orion Bar PDR \citep{2023ApJ...951...97L}, irradiated by the Trapezium cluster at $\sim10^5~L_{\odot}$ (estimated using an effective temperature of 39000 K, and effective radius $10.6~R_{\odot}$ \cite{2006A&A...448..351S}) found that the radiation field was not the preferred alignment axis, and that grains large enough for k-RAT are destroyed by rotational disruption (RATD), suggesting k-RAT is self-limiting in strong radiation environments. Since the luminosity of DR21(OH) does not exceed $10^5~L_{\odot}$ \citep{Harvey1977,Mangum1991,Jakob2007}, k-RAT is unlikely to dominate grain alignment in our region.

\subsection{Local enhancement of grain alignment efficiency from embedded protostars}
\begin{figure*}[htb!]
\centering
\includegraphics[width=0.9\linewidth]
{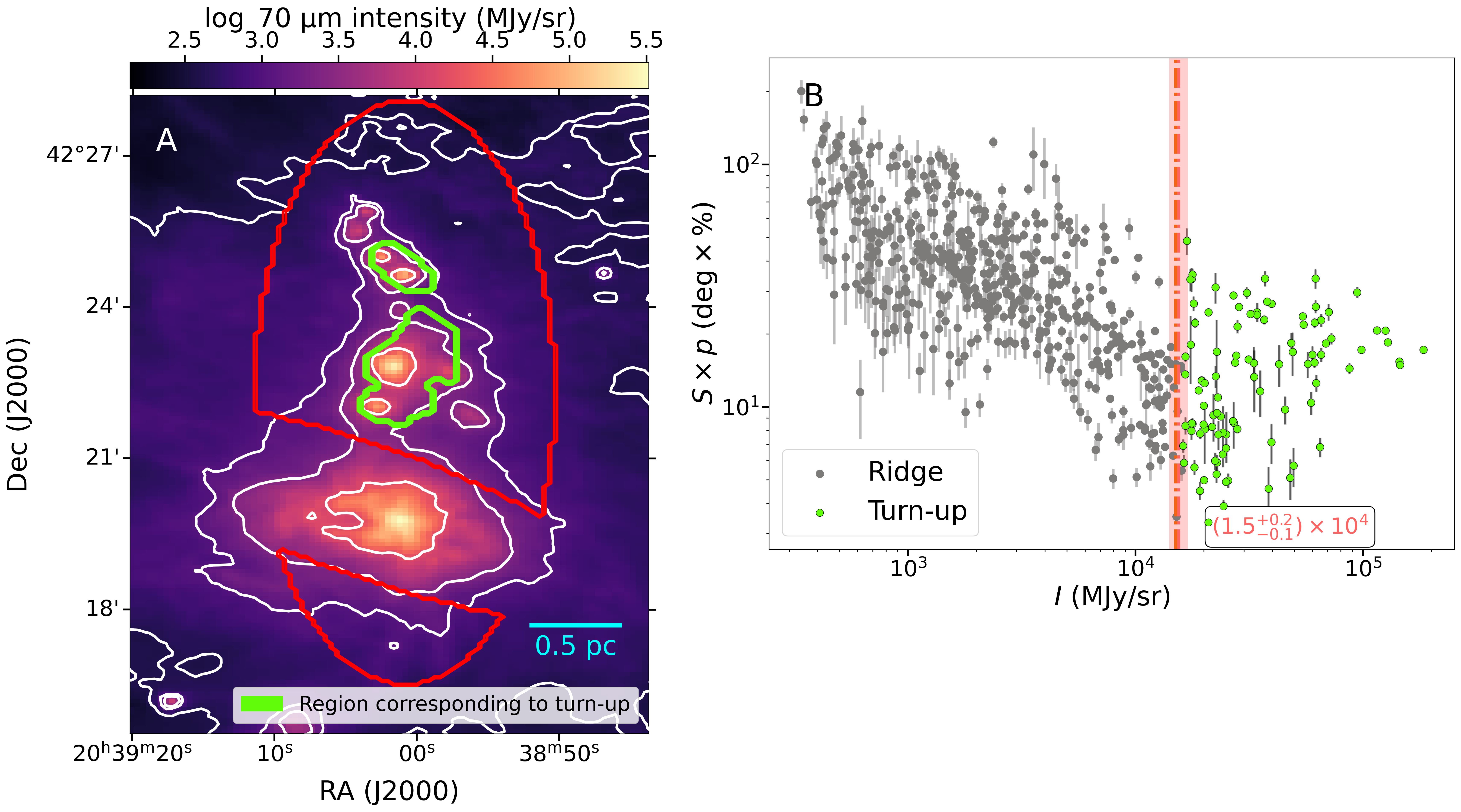}
  \caption{Panel A (left) displays the {\it Herschel}/PACS 70 \um{} intensity map of DR21 with the Ridge outlined in red. Contours are plotted at levels of [2.61, 3.05, 3.40, 4.21] (log MJy/sr), corresponding to the 50th, 90th, 95th, and 99th pixel value histogram percentiles, respectively. The light-green outline indicates the spatial extent where the increase in grain alignment efficiency is observed. Panel B (right) shows $S \times p$ versus $I$ for the Ridge, with data points in gray. Data points exhibiting increased grain alignment efficiency are shown in light green. The transition is marked by a red dot-dashed line at $1.5^{+0.2}_{-0.1} \times 10^4$\,MJy/sr, with the $\pm1\sigma$ uncertainty in the break location shown as red shading.}

  \label{fig:70muananlysis_stokesI}
\end{figure*}

RAT alignment theory predicts that grain alignment efficiency increases towards protostars due to enhanced local radiation fields, resulting in higher polarization fractions near emission peaks. This prediction was explored analytically in \cite{Hoang_Tram_Lee_Diep_Ngoc_2021}, who proposed that the enhanced alignment would appear as a flattening of the $p$–$I$ relation at the highest intensities in the absence of RAT-D. 

Although a flattening in the $p$–$I$ relation has not been reported, previous observational studies have attributed shallower $p$ versus $I$ slopes to this enhancement of RAT alignment from embedded radiative sources. For example, \cite{2020NatAs...4.1195P}'s work on Serpens South and \citep{Ching_Qiu_Li_Ren_Lai_Berry_Pattle_Furuya_Ward-Thompson_Johnstone_et}'s work on DR21, both of which reported shallower slopes. In contrast, we detect a clear flattening, with the polarization fraction initially following a negative slope of $\alpha = -0.45\pm0.03$ but flattening to $\alpha =0.14\pm0.11$, beyond $I = 1.6^{+0.3}_{-0.3} \times 10^4$ MJy/sr (panel A of Figure \ref{fig:p_vs_S_and_I}).This behaviour is consistent with the prediction of \cite{Hoang_Tram_Lee_Diep_Ngoc_2021}. 

Previously, a direct observational signature of the prediction was reported in \cite{2020A&A...644A..11L}, which studied polarized emission from several stacked ALMA cores and observed a flat $S\times p$ as a function of normalized $\rm N_{H_2}$  (their Figure 10) for high luminosity cores ($>10~L_{\odot}$, which they note was an arbitrary choice). In comparison, we observe a turn-up in the behaviour of $S\times p$ (panel C of Figure \ref{fig:s_vs_I_sp_vs_I}) at the highest intensities, with the slope changing from a negative to a positive value at $I = 1.5^{+0.2}_{-0.1} \times 10^4$ MJy/sr, reaching $\alpha = 0.37\pm0.13$. This difference may arise from the much smaller spatial scales ($<10^{-4}$~pc) probed by ALMA, as well as from the effect of averaging measurements of several proto-stellar cores that may themselves span a range of local physical conditions. In addition, the ALMA cores analyzed in that study are predominantly low-mass protostars, whereas the DR21 region hosts high-mass protostars.

Altogether, our results suggest that grain alignment is enhanced by internal anisotropic radiation from proto-stellar cores, a hypothesis that is further tested in the following sections.

\subsection{The effects of protostellar cores on grain alignment efficiency}

To test the interpretation of local enhancement of grain alignment originating from embedded protostars, a 70 \um{} emission map of the region was examined. If the local enhancement of grain alignment originates from embedded protostars, it is expected to spatially coincide with regions where dust is heated by these internal radiation sources. Although the \textit{Herschel}-based dust temperature map could, in principle, be used for this purpose, its applicability was limited because a significant portion of the map near DR21(OH) was blanked out due to saturation caused by the bright cores \citep{Pokhrel2020}. Instead, the 70 \um{} emission was used as a tracer of local internal radiation from embedded protostars \citep{2010ApJ...725..677L}. The map, obtained from the Herschel Science Archive was observed with the {\it Herschel}/PACS instrument \citep{2010A&A...518L...2P} and regridded to match the resolution and spatial extent of the \sofia{}/\hawc{} maps and is shown in Panel A of Figure \ref{fig:70muananlysis_stokesI}. 

The correlation between enhanced grain alignment efficiency and high 70 \um{} intensity was examined as follows. The broken power-law fit to $S\times p$ versus $I$ identified a transition point at $1.5^{+0.2}_{-0.1}\times10^{4}$ MJy/sr, above which the slope changes from negative to positive. Data points corresponding to this "turn-up" are highlighted in light-green in Panel B of Figure \ref{fig:70muananlysis_stokesI}. Based on the sky positions of these points, a light-green boundary was drawn on the 70 \um{} emission map encompassing the regions corresponding to the light-green turn-up data points. The region inside the light-green boundary contains areas of high 70 \um{} emission, as seen from the contours plotted in Panel A of Figure \ref{fig:70muananlysis_stokesI}, indicating the presence of strong local radiation fields originating from embedded proto-stellar cores.

The observed correlation between strong 70 \um{} emission and increased $S\times p$ supports the hypothesis that embedded protostars enhance grain alignment in denser, shielded regions. The 70 \um{} analysis provides evidence that the observed increase in grain alignment efficiency at high intensity is a real physical effect driven by internal protostellar radiation.

\subsection{Polarization radiative transfer model}
\label{sec:modeling}

To further test the interpretations of the observed polarization trends, a simplified radiative transfer model of a spherical proto-stellar envelope with a central luminous source was implemented using the $\mathrm{DustPOL\_py}$\footnote{Github: https://github.com/lengoctram/$\mathrm{DustPOL\_py}$}\citep{2025A&A...703A.192T}.
The purpose of this model was to illustrate whether a minimal set of
assumptions could reproduce the observed behaviour of the polarization fraction at moderate to high intensities.
Based on our analysis of the \textit{Herschel} column density data (Paper I), a cylindrically  symmetric envelope described by a Plummer-like density distribution was adopted:
\begin{equation}
\hspace{75pt} n(r) = \frac{n_0}{\left[1 + (r/r_\mathrm{flat})^2\right]^{p/2}},
\end{equation}
with central density $n_{\rm H} = 6 \times 10^{6} \, \mathrm{cm}^{-3}$, flat radius $r_\mathrm{flat} = 0.1 \, \mathrm{pc}$ and power-law index $p = 2.4$. The outer radius was set to $r_{\rm out} = 1\,\mathrm{pc}$, corresponding to $10\,r_{\rm in}$, to match the extent implied by the observed column-density structure.
A spherical core was assumed to be heated by a single embedded source with luminosity $L_\star = 10^5 \, L_\odot$, representing a high-mass protostar/cluster \citep{Harvey1977}. The central source was modeled as a blackbody and the radiation field was computed wavelength-by-wavelength including extinction \citep[see Equations 4-5][]{2026arXiv260100731T}. The spectral shape of the illuminating source is therefore self-consistently accounted for in the model. A standard interstellar radiation field (Habing Field = 1) enveloping the core was also included. The internal radiation propagated outward through the dusty envelope, attenuated primarily within $r_\mathrm{flat}$. For comparison, \cite{Hoang_Tram_Lee_Diep_Ngoc_2021} adopted a much smaller flat radius ($r_\mathrm{flat} = 500 \, \mathrm{AU}$), leading to significantly less attenuation and a more extended RATD effect. As the \sofia{}/\hawc{} beam is much larger than the region over which RATD is expected to dominate \citep[$\sim1000$ AU,][]{Giang_Le}, the effects of RATD could not be resolved with our data and model.

The dust grains were assumed to be a mixture of material \textsc{astrodust} \citep{2023ApJ...948...55H}, following an MRN-like size distribution $n(a) \propto a^{-3.5}$ with a minimum grain size $a_\mathrm{min} = 3~\mathring{A}$ and a maximum size $a_\mathrm{max}$ ranging from $0.5$ to 3 \um{}. Oblate grains with an aspect ratio of 1.4 were adopted, consistent with previous polarization modeling studies \citep{Hensley2023}. Grain alignment was treated using the RAT alignment theory \citep{2007MNRAS.378..910L,Hoang2014}, including the effects of RATD using the \textsc{DustPOL\_py} model under the GRADE-POL framework (\citealt{Tram2025arXiv}). In this model, a uniform magnetic field and perfect alignment efficiency by RAT ($f_{\mathrm{max}}=1$, $f_{max}$ is the alignment function (see Equation 16 in \citealt{Tram2025arXiv}) were assumed. In this framework, the minimum alignment size $a_\mathrm{align}$ was computed, smaller than which grains do not align effectively with the magnetic field. Lower values of $a_\mathrm{align}$ correspond to a greater number of aligned grains and thus
higher expected polarization fractions.

\begin{figure}[htb!]
    \centering
    \includegraphics[width=\columnwidth]{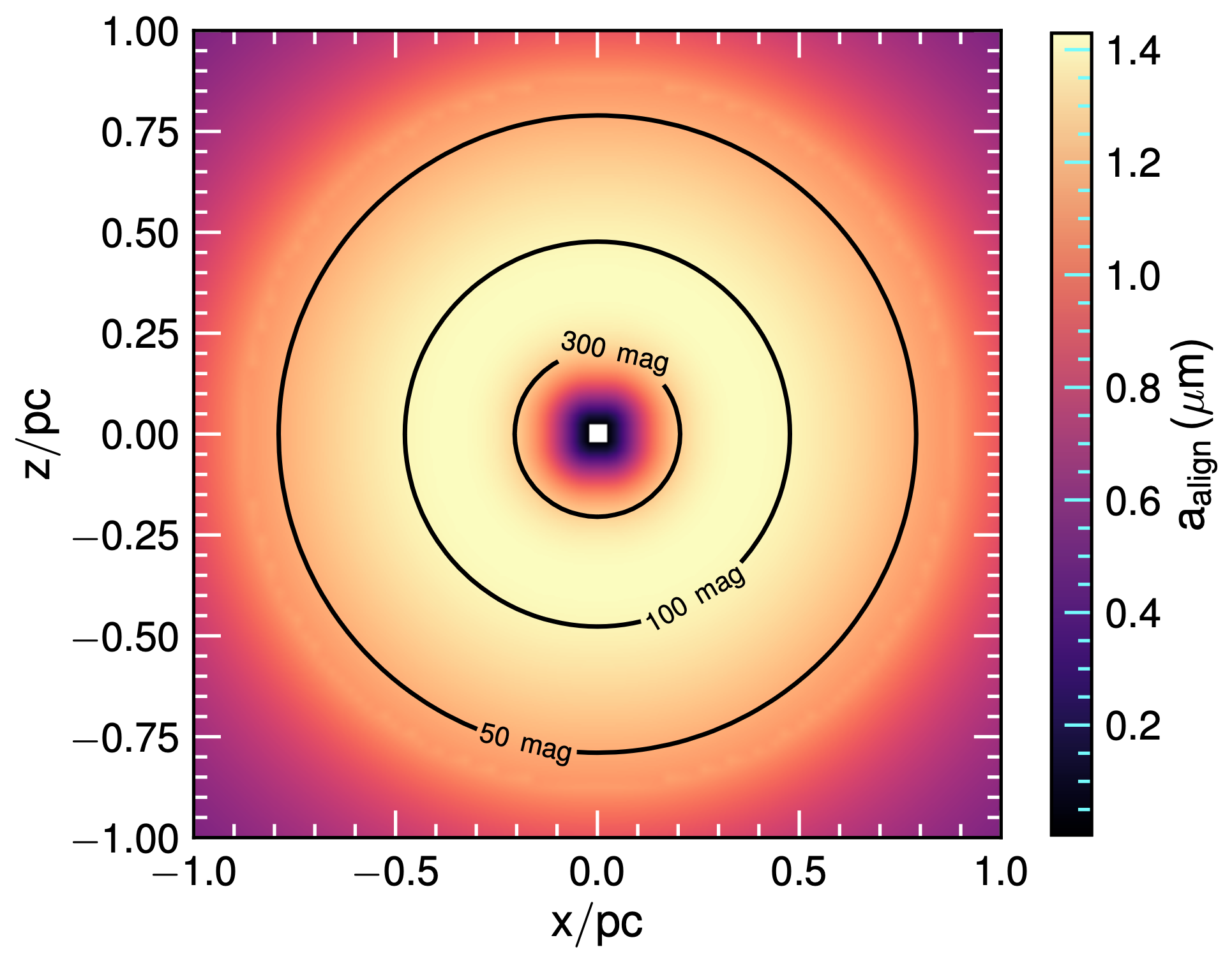}
    \caption{The central slice of the three-dimensional $a_{\rm align}$ cube in the $x$–$z$ plane. The color scale indicates $a_{\rm align}$ in \um{}, with lower values corresponding to a higher number of aligned grains. Black contours denote lines of constant visual extinction at 50, 100, and 300 mag. The central white square marks the location of the embedded radiation source.}
    \label{fig:alignmap}
\end{figure}

Figure \ref{fig:alignmap} shows the x-z cut of the $a_\mathrm{align}$ map with contours of visual extinction ($A_V$) overlaid. The map shows that $a_\mathrm{align}$ increases inward from the outer envelope, indicating a reduced alignment efficiency and lower polarization fractions. However, once the extinction exceeds $A_V \sim 300$ mag, $a_\mathrm{align}$ decreases sharply, signaling a reactivation of alignment due to internal radiative feedback. When comparing the scales of alignment reactivation, the model predicts a reactivation of alignment at scales of $\approx0.4~pc$ (Figure~\ref{fig:alignmap}), whereas Figure~\ref{fig:70muananlysis_stokesI} indicates a scale of $\approx0.6~pc$. Although the model predicts a smaller scale, the two scales remain comparable given the simplifying assumptions in the modeling.

\begin{figure*}[htb!]
\centering
\includegraphics[scale = 0.5]{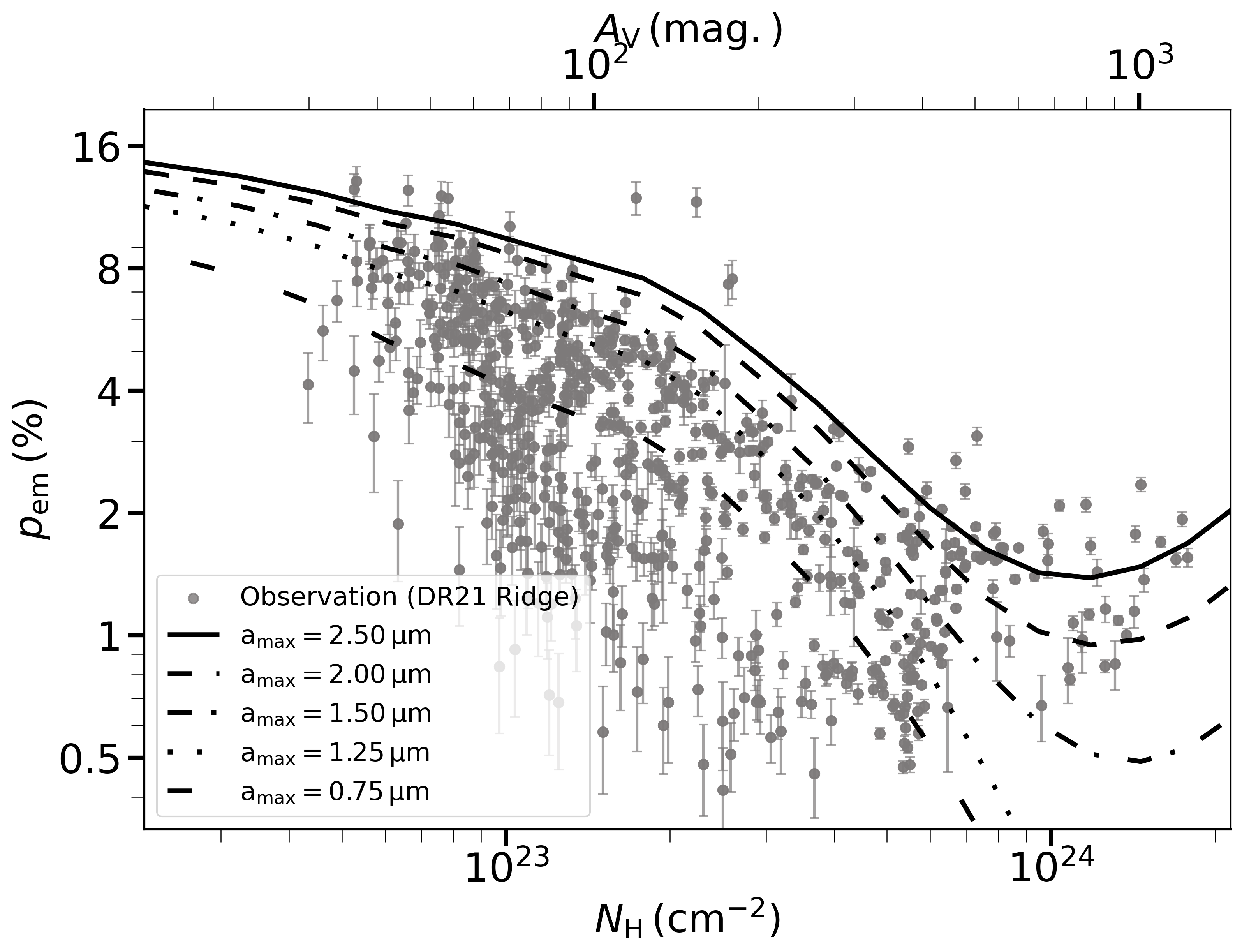}
\caption{Comparison between the radiative transfer model predictions and the observed polarization fraction versus hydrogen column density (N$_\mathrm H$) for the DR21 Ridge. Luminosity of the central radiating source was set to $L = 10^5~L_{\odot}$. The x-axis on top shows the $A_V$ (mag) extinction values. Each model curve corresponds to a different maximum grain sizes in the 0.75 - 2.5 \um{} range. All models with $\mathrm{a_{max}} > 1.25$ \um{} reproduce the observed behaviour at high (N$_\mathrm H$) values (several $> 10^{23}$ cm$^{-2}$), including the turn-up region in the \sofia{}/\hawc{} observations. \label{fig:pfracmodel}}
\end{figure*}

Figure~\ref{fig:pfracmodel} shows the resultant model polarization fraction as a function of column density where the gray circles represent the \sofia{} data points. The figure implies that polarization fractions decrease with increasing column density to $A_V \approx 300$, and then rise again at
higher densities. In this study, the threshold for the change in trend was observed at $I = (1.6 \pm 0.3) \times 10^4$~MJy\,sr$^{-1}$.  The corresponding $\rm N_{{H}_2}$ was computed as the mean  $\rm N_{{H}_2}$ evaluated within the intensity range $(1.6 \pm 0.3) \times 10^4$~MJy\,sr$^{-1}$. This value of column density was then converted to $A_V$ using the $\rm N_{{H}_2}$ map from \citet{Pokhrel2020} and the relation $\rm N_{H}/E(B-V) = 5.8 \times 10^{21}~\mathrm{cm}^{-2}~\mathrm{mag}^{-1}$ \citep{Bohlin1978}. For molecular gas in dense cloud environments with $R_V = 4$--$5$, this gives $A_V/\mathrm{N_{H_2}} = (1.38$--$1.72) \times 10^{-21}~\mathrm{mag~cm}^{2}$, yielding $A_V \approx 290$--$360$ mag.

The point where the polarization fraction increases is farther from the center (deeper inside the envelope) for large grains (see Figure \ref{fig:pfracmodel}). For smaller grains, the $a_{align}$ location is much closer to the central source. This is because longer wavelengths of radiation from the internal source can penetrate farther from the source (shorter wavelengths become extincted more strongly). The turn-up was observed for all models having $\mathrm{a_{max}}>1.25$ \um{}. Thus, in addition to the absolute value of the polarization fraction, the turn-up also provides constraints on grain sizes. For a lower stellar luminosity, e.g., $L_\star = 10^4 \, L_\odot$ (see Appendix \ref{sec:diff_L}), the turn-up was not seen because the internal radiation field is not sufficient to align grains in the envelope. 

It should be noted that the adopted radiative transfer setup involves simplifying assumptions, notably the use of a single embedded protostar. However, the region contains multiple massive protostars  \citep{2010A&A...524A..18B, Beerer2010, Cheng2022}, including DR21(OH), which likely produce a more complex radiation field than assumed.

\subsection{Comparison of depolarization in sub-filaments and Ridge}

In this study, the dust polarization in the sub-filaments was also analyzed. A comparison between the Ridge and the sub-filaments, indicates that the overall patterns of dust grain alignment are broadly similar. For the $p$ versus $I$ relation, the slope in the sub-filaments is steeper than that of the Ridge ($I < 1.6^{+0.3}_{-0.3}$ MJy/sr). The relatively shallow slope in the Ridge (see panel A and B of Figure \ref{fig:p_vs_S_and_I}) can be attributed to the presence of several embedded protostars \citep{2010A&A...524A..18B,2010A&A...520A..49S}, which are absent in the sub-filaments. 

In panels C and D of Figure \ref{fig:p_vs_S_and_I}, the $p$ versus $S$ relation reveals a steeper slope in the sub-filaments, indicating a stronger depolarization caused by magnetic field tangling. When making this comparison, the different dynamic ranges in Stokes $I$ sampled by these regions must be taken into account. From the $p$ versus $S$ plots, the relation for the Ridge appears more scattered, likely as it samples data across a much wider dynamic range in Stokes $I$, spanning over three orders of magnitude. This wide range in $I$ is compressed in the $p$ versus $S$ plane, where each $S$ bin contains regions with very different $I$ values and consequently distinct polarization fractions. In contrast, the sub-filaments span only about one order of magnitude in intensity ($\sim10^2$–$10^3$ MJy sr$^{-1}$). This results in a more uniform sampling and thus giving a clearer and steeper trend in $p$ versus $S$.

The angular dispersion function, $S$, remains nearly constant with intensity in both environments (see panels A and B of Figure \ref{fig:s_vs_I_sp_vs_I}). Comparing the $S\times p$ trends, the sub-filaments exhibit a similar $S\times p$ versus $I$ slope to the ridge over the shared intensity range (below $4.7^{+0.9}_{-0.8} \times 10^3$ MJy/sr). The sub-filaments show no RAT re-activation, as expected given the lack of internal radiation sources at these intensity levels.

\subsection{Comparisons with previous work on DR21}
The two most intense star forming sites in DR21, DR21(M) and DR21(OH) have been the focus of mm/submm polarimetric observations for over three decades. The early 800 \um{} polarimetric study of DR21(M) \citep{1994A&A...286..579M} and near-IR H$_2$ imaging polarimetry \citep{1999MNRAS.304..406I} found a decreasing polarization fraction, a "polarization hole", towards the peak position, but did not have sufficient sensitivity to characterize the slope of polarization fraction vs emission intensity or $A_V$. Subsequent work at 350 \um{}, again towards DR21(M) reinforced this polarization hole phenomenon but did not explore the $p-I$ relation \citep{2009ApJ...694.1056K}. As these studies concentrated on DR21(M), which is masked in the present analysis, a direct comparison is not possible.  

The $p-I$ relation throughout the DR21 region was studied by \cite{2022ApJ...941..122C}. Conducted as part of the BISTRO survey using the POL-2 polarimeter and the SCUBA-2 camera on the James Clerk Maxwell Telescope, the \cite{2022ApJ...941..122C} observations targeted the DR21 filament at 850 \um{} and 14.1'' resolution. They reported a polarization fraction-intensity slope of $\alpha = -0.34\pm0.05$ for the region around DR21(OH), called the "north filament" in their study (their Figure 4). This value is identical to the slope of $\alpha = -0.35\pm0.02$ reported in this study. Similar to our study, the shallow slope was attributed to grains remaining aligned in the dense cores due to anisotropic radiation from embedded protostars; however, this was not further explored. 

A direct comparison of polarization fractions between the two studies is not straightforward for several reasons. Differences in wavelength imply that distinct/different grain populations may dominate the emission, and RAT efficiency is expected to vary accordingly. The shorter wavelength (214 \um{}) observations are more sensitive to warmer dust than the JCMT 850 \um{} data, a distinction particularly relevant for the high intensity, warm peaks of DR21(OH) \citep{2012A&A...543L...3H}. Finally, sky subtraction for both instruments is a significant source of uncertainty for polarization-fraction studies, especially at low intensities. Despite the difficulty in direct comparison, it would be of interest to repeat the analysis of the broken power law in $p$ vs $I$ and $S\times p$ vs $I$, presented here, using the \cite{2022ApJ...941..122C} data, to investigate whether these trends are also consistent.

\section{Conclusions}\label{sec:conclusions}
We tested predictions from RAT theory for dust grain alignment in the star-forming region DR21 using thermal dust emission polarimetry observed using the \hawc{} instrument at 214~\um{} on \sofia{} as part of the SIMPLIFI project. Our main findings are summarized below:

\begin{enumerate}
\item We recovered significant polarization intensity even at low intensities, indicating that \sofia{}/\hawc{} can effectively trace magnetic fields over a wide range of intensities ($10^3-10^5$ MJy/sr).
    
\item The polarization fraction ($p$) decreases with increasing intensity $I$, similar to previous studies of dense cores. However, at the highest intensities (> $1.6^{+0.3}_{-0.3} \times 10^4$ MJy/sr), the trend flattens, suggesting deviation from standard depolarization behaviour. 
    
\item To understand the behaviour of $p$-$I$, we examined magnetic turbulence (via angular dispersion, $S$) and grain alignment efficiency (via $S \times p$) as functions of intensity. No significant correlation between angular dispersion and intensity was found, implying magnetic turbulence levels remain similar across the probed scales. In contrast, grain alignment efficiency increases at the highest intensities (> $1.5^{+0.2}_{-0.1} \times 10^4$ MJy/sr), suggesting enhanced grain alignment in regions surrounding embedded sources.

\item To test the hypothesis of enhanced grain alignment, {\it Herschel}/PACS the 70 \um{} intensity map was examined. The regions of enhanced alignment efficiency were found to spatially coincide with strong 70 \um{} emission, supporting the hypothesis that anisotropic radiation from protostars drives the reactivation of RAT alignment.    

\item A simple analytical model of a centrally heated envelope with an embedded luminous protostar was implemented to test the results of this study. The model reproduces the observed behaviour of an initial decrease in polarization fraction as function of column density followed by a flattening at highest extinctions, further supporting the interpretation of internally powered RAT alignment. 
\end{enumerate}

Given the scarcity of direct observational evidence for locally enhanced polarization associated with embedded protostars, systematic searches for similar signatures of RAT enhancement in other star-forming regions are essential. Such efforts will enable a more comprehensive assessment of the physical conditions under which radiative torques can effectively enhance grain alignment and shape the observed polarization properties.

\begin{acknowledgements}
This paper is based on observations made with the NASA/DLR Stratospheric  Observatory for Infrared Astronomy (SOFIA). SOFIA was jointly operated by the Universities Space Research Association, Inc. (USRA), under NASA contract  NNA17BF53C, and the Deutsches SOFIA Institut (DSI) under DLR contract 50 OK 0901 to the University of Stuttgart. TGSP gratefully acknowledges support by NASA award \#09-0215 issued by USRA and awards from National Science Foundation under grant no. AST-2009842 and AST-2108989. DS acknowledges support of the Bonn-Cologne Graduate School, which is funded through the German Excellence Initiative as well as funding by the Deutsche Forschungsgemeinschaft (DFG) via the Collaborative Research Center SFB 1601 ‘Habitats of Massive Stars Across Cosmic Time’ (subprojects B1 and B4). This research was carried out in part at the Jet Propulsion Laboratory, which is operated by the California Institute of Technology under a contract with the National Aeronautics and Space Administration (80NM0018D0004).  GP and SK acknowledge support from the Knut and Alice Wallenberg Foundation Fellowship program under grant number 2023.0080. 
We thank R. Gutermuth and R. Pokhrel for kindly providing us the Herschel based column density and temperature maps from \citet{Pokhrel2020}.
\end{acknowledgements}

\bibliographystyle{aa} 
\bibliography{sample} 
\begin{appendix}

\section{Bias in the detected slope of $p$-$I$ relation}{\label{sec:bias_in_alpha}}
In \cite{2019ApJ...880...27P}, the authors introduced a parameter $\sigma_{QU}$ with the units of intensity and a single value representing the rms noise in measurements of Stokes $Q$ and $U$ calculated using

\begin{equation}
\hspace{65pt}\langle \sigma_{QU}\rangle = \frac{1}{2N}\sum_{i=1}^N\left(\sqrt{V_{Q,i}} + \sqrt{V_{U,i}}\right)
\end{equation}

where $V_Q$ and $V_U$ are the variance values associated with each stokes $Q$ and $U$ pixel and N is the number of pixels in the data set. Using this parameter, the authors constructed the following model to characterize the observed polarization fraction:

\begin{equation}
\hspace{85pt}p(I) = p_0\left(\frac{I}{I_0} \right)^{\alpha},
\end{equation}
 Here, $I_0$ is taken to be equal to $\sigma_{QU}$ (i.e, $I_0 = \sigma_{QU}$) under the assumption that the power-law relation between $I$ and $p$, observed intensity and polarization fraction, was applicable to all measurements above the noise level of the data. Similarly, $p_0$ represents the polarization fraction at the noise level: $p_0=p(I=\sigma_{QU})\equiv p_{\sigma_{QU}}$.

The authors demonstrated that in the low S/N regime, $p \ll \sigma_p$; ($\sigma_p = \sigma_{QU}/I$), an artificial slope of $\alpha = -1$ dominated the polarization-intensity relation regardless of the true value of $\alpha$. To quantify this bias the parameter $\left(I/\sigma_{QU}\right)$ was introduced. A critical threshold  for the parameter was identified below which the estimates of $\alpha$ become biased toward the low S/N $\alpha = -1$ regime. This critical value is given by:

\begin{equation}
\hspace{75pt}\left(I/\sigma_{QU}\right)_{crit} = \left(\frac{1}{p_{\sigma_{QU}}}\sqrt{\frac{\pi}{2}}\right)^{\frac{1}{1-\alpha}},
\end{equation}

Figure 1 of \cite{2019ApJ...880...27P} shows how $\left(I/\sigma_{QU}\right)_{crit}$ varies with $\alpha$ and $p_{\sigma_{QU}}$, demonstrating that reliably characterizing power-law behaviour shallower than $\alpha = 1$ requires a sufficient number of data points with $I/\sigma_{QU} > \left(I/\sigma_{QU}\right)_{\mathrm{crit}}$. For the case of our dataset, a direct estimation of $\left(I/\sigma_{QU}\right)_{\mathrm{crit}}$ was not possible because Stokes $I$ values were unavailable at levels comparable to $\sigma_{QU}$, preventing determination of $p_{\sigma_{QU}}$. However, our data show a minimum value of $I/\sigma_{QU} = 10$, indicating all measurements exceed at least this threshold. According to their Figure 1, this ensures reliable recovery of $\alpha$ for most combinations of true $\alpha$ and $p_{\sigma_{QU}}$. In addition to this, since low S/N bias would steepen the recovered slope, our relatively shallow measured slope suggests that our $\alpha$ estimates are not significantly affected by such bias.

\section{Trends of $p$ and dust grain alignment vs $N_{\rm H_2}$}{\label{sec:app_vsN}}

\begin{figure*}[htb!]
\centering
\includegraphics[width=1\linewidth]{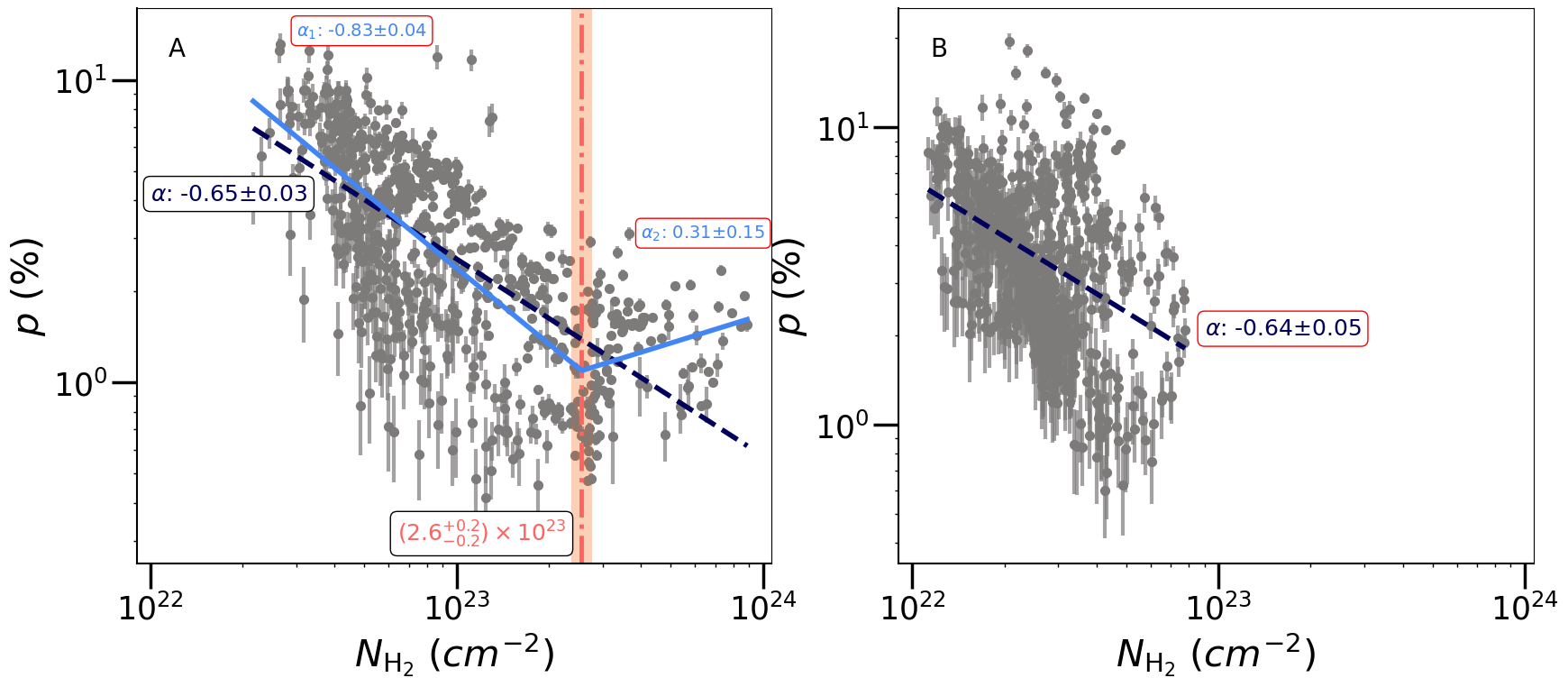}
\caption{Trends of $p$ vs $N_{\rm H_2}$. The data points are plotted in gray with the Ridge in panel A and the sub-filaments in panel B. The single power-law fits are shown as a dashed dark blue line in both panels. For the Ridge (panel A), a broken power-law fit is also included. The break is marked by vertical red dot-dashed line, with the error in estimation of the break point shown as red shading ($\pm1\sigma$) around the vertical line. \label{fig:p_vs_N}}
\end{figure*}

\begin{figure*}[htb!]
\centering
\includegraphics[width=1\linewidth]{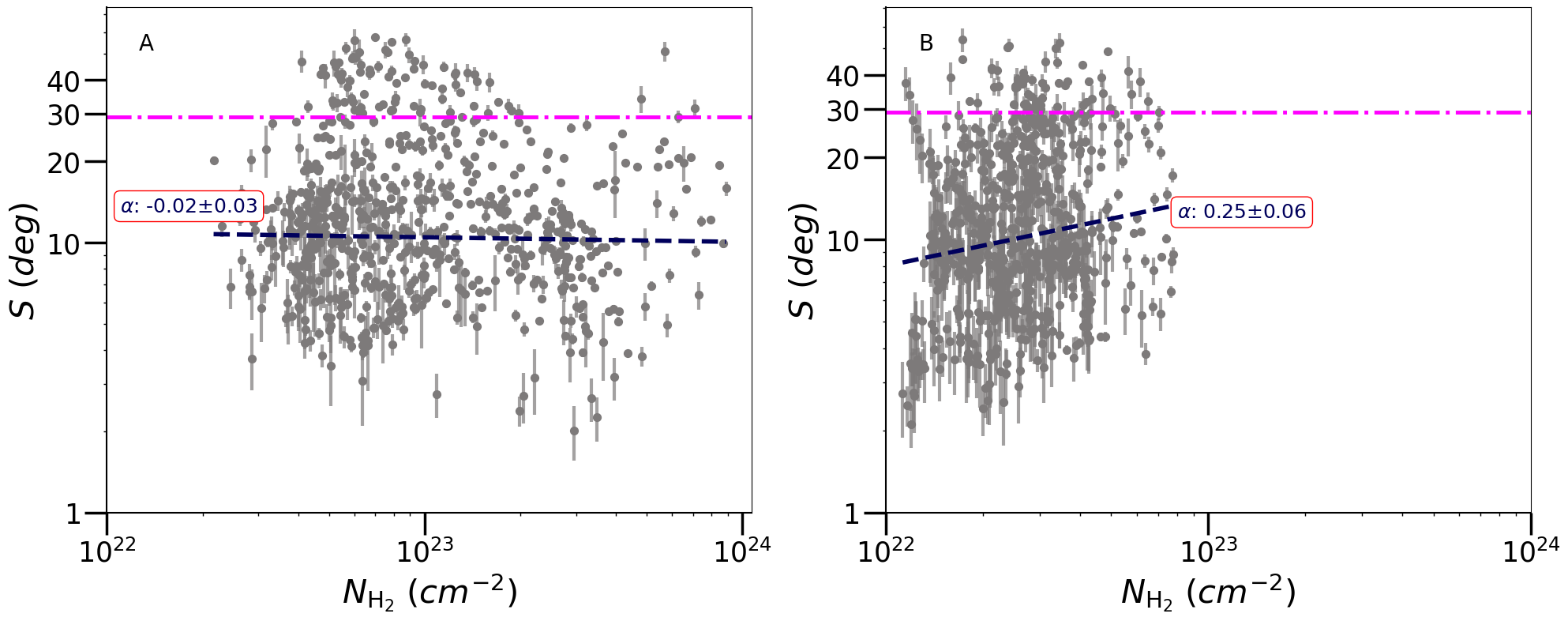}
\caption{As in Figure \ref{fig:p_vs_N} but showing the relationship between $S$ and $N_{\rm H_2}$. The saturation limit of angular dispersion is marked by a magenta dot-dashed line at $29.2^\circ$.  \label{fig:s_vs_N}}
\end{figure*}

\begin{figure*}[htb!]
\centering
\includegraphics[width=1\linewidth]{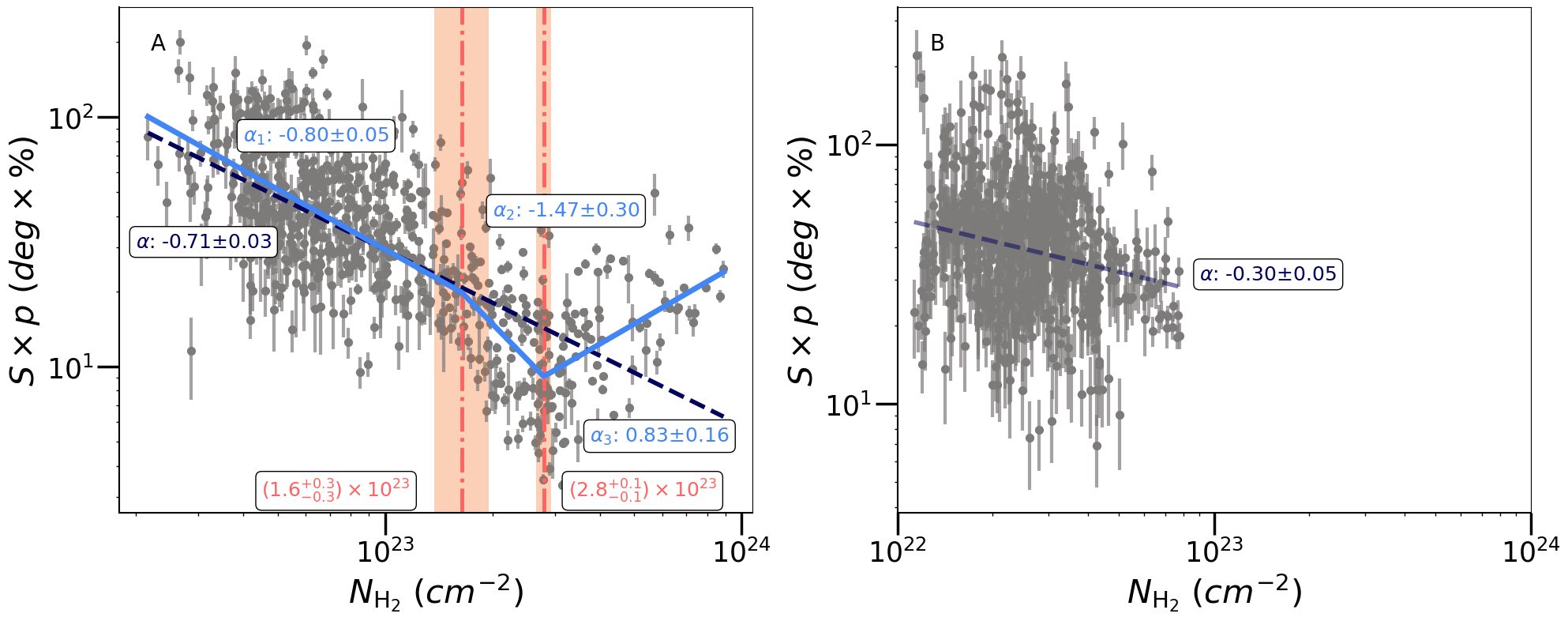}
  \caption{As in Figure \ref{fig:p_vs_N} but for the relationship between $S\times p$ and $N_{\rm H_2}$ with the Ridge shown in panel A and the sub~-~filament in panel B. For the Ridge, a power law with two break breaks was fitted. The breaks are marked by vertical red dot-dashed lines, with the error in the estimation of break points shown as red shading ($\pm1\sigma$) around the vertical lines.\label{fig:sp_vs_N}}
\end{figure*}

To examine whether the trends in Section 4 change when using $N_{H_2}$ instead of intensity and to provide a complementary analysis to the intensity based analysis, the relationship between the polarization fraction ($p$) and the column density ($N_{\rm H_2}$) was examined. 

Before interpreting these results, it is important to note the assumptions underlying the derivation of the column density maps. A fixed dust emissivity index ($\beta = 1.5$) was adopted, which may not hold uniformly across the region. Furthermore, a significant portion of the temperature map was blanked out in the densest areas \citep{Pokhrel2020}, introducing additional uncertainty into the derived column densities.

Figure~\ref{fig:p_vs_N} shows the $p$–$N_{\rm H_2}$ relation for the Ridge (left) and the sub-filaments (right). As expected, both regions exhibit a clear anti-correlation, similar to the case for intensity. A power-law relation (Equation~(\ref{eq:singlefit})) fit to the data using the statistical methods described in Section~\ref{sec:method_stat}, yielding $p \propto N_{\rm H_2}^{-0.65\pm0.03}$ ($\rho = -0.78$, p-value $\ll0.001$) for the Ridge and $p \propto N_{\rm H_2}^{-0.64\pm0.05}$ (r = -0.35, p-value $\ll$ 0.001) for the sub-filaments. Both the Ridge and sub-filaments have the same slope indicating similar levels of dust grain alignment in both regions. BIC comparisons showed a two component broken power-law model was the best fit for the Ridge, while a single power-law sufficed for the sub-filaments. The fit yielded a break at $N_{\rm H_2} = 2.6^{+0.2}_{-0.2} \times 10^{23}$cm$^{-2}$. The slope is steep and negative ($\alpha \approx -0.83\pm0.04$) below the break, while the trend flattens ($\alpha \approx0.31\pm0.11$) above it, indicating a near-constant polarization fraction as a function of column density at higher densities. 

Although the values of the slopes are quantitatively different than those found in Section 4, the overall qualitative behaviour of $p$ vs $N_{H_2}$ is similar. For the case of a single power law fit, the slopes were found to be similar for the Ridge and sub-filament. But when the slopes at similar column densities are compared (excluding  data points with $N_{\rm H_2} > 2.6^{+0.2}_{-0.2} \times 10^{23}$cm$^{-2}$), the Ridge exhibits a significantly greater negative slope compared to the sub-filament. 
As shown in \citet{2019ApJ...880...27P}, the slope of the $p-N_{\rm H_2}$ relation is generally shallower than that of the $p$ versus $I$ relation. However, in this study an opposite trend was observed. This difference might arise from the assumptions used in constructing the different column density maps. For example, the adoption of a fixed emissivity index and temperature saturation in the densest regions can introduce systematic biases in the inferred $p-N_{\rm H_2}$ slope.

 Figure \ref{fig:s_vs_N}, compares the values of angular dispersion ($S$) and column density ($N_{\rm H_2}$) for the Ridge on the left and the sub-filaments on the right. A power-law fit to the data using Equation (\ref{eq:singlefit}) yields $S\propto N_{\rm H_2}^{-0.02\pm0.03}$ ($\rho = -0.02,$ p-value$= 0.58$) for the Ridge and $S\propto N_{\rm H_2}^{-0.25\pm0.06}$ ($\rho = 0.14,$ p-value$= 0.005$) for the sub-filament, indicating minimal correlation in the Ridge but a significant correlation in the sub-filaments.

Figure \ref{fig:sp_vs_N} compares the values of grain alignment efficiency ($S \times p$) and $N_{\rm H_2}$ for the Ridge on the left and the sub-filament on the right. A negative correlation is observed between dust grain alignment efficiency and column density. A power-law fit to the data yields $S\times p\propto N_{\rm H_2}^{-0.71\pm0.03}$ ($\rho = -0.76,$ p-value $\ll 0.001$) for the Ridge and $S\times p\propto-N_{\rm H_2}^{-0.30\pm0.05}$ for sub-filament ($\rho = -0.21,$ p-value $\ll 0.001$). In addition to a single power law fit, we fitted broken power-law models with one, two, and three breaks. The BIC comparison favored a two-break model for the Ridge, with breaks at $N_{\rm H_2} = 1.6^{+0.3}_{-0.3}\times10^{23}\mathrm{cm^{-3}}$ and $2.8^{+0.1}_{-0.1}\times10^{23}\mathrm{cm^{-3}}$. In the first segment ($N_{\rm H_2} < 1.6^{+0.3}_{-0.3}\times10^{23}\mathrm{cm^{-3}}$), the slope is $\alpha \approx -0.80\pm0.05$. Between $1.6^{+0.3}_{-0.3}\times10^{23}$ and $2.8^{+0.1}_{-0.1}\times10^{23}\mathrm{cm^{-3}}$, the negative slope becomes steeper, with $\alpha \approx -1.47\pm0.30$. Beyond $2.8^{+0.1}_{-0.1}\times10^{23}\mathrm{cm^{-3}}$, the trend reverses, and the slope becomes positive with $\alpha \approx 0.83\pm0.16$. The necessity for two break points and the rise at high column densities is consistent with the $S\times p$ versus $I$ results shown in Section \ref{sec:results}. We conclude that despite the uncertainties associated with the $N_{H_2}$ map, our main results from the analysis using $I$, remain when using column density.

\section{Effect of embedded proto-stellar luminosity on grain alignment}\label{sec:diff_L}

\begin{figure*}[htb!]
\centering
\includegraphics[scale = 0.5]{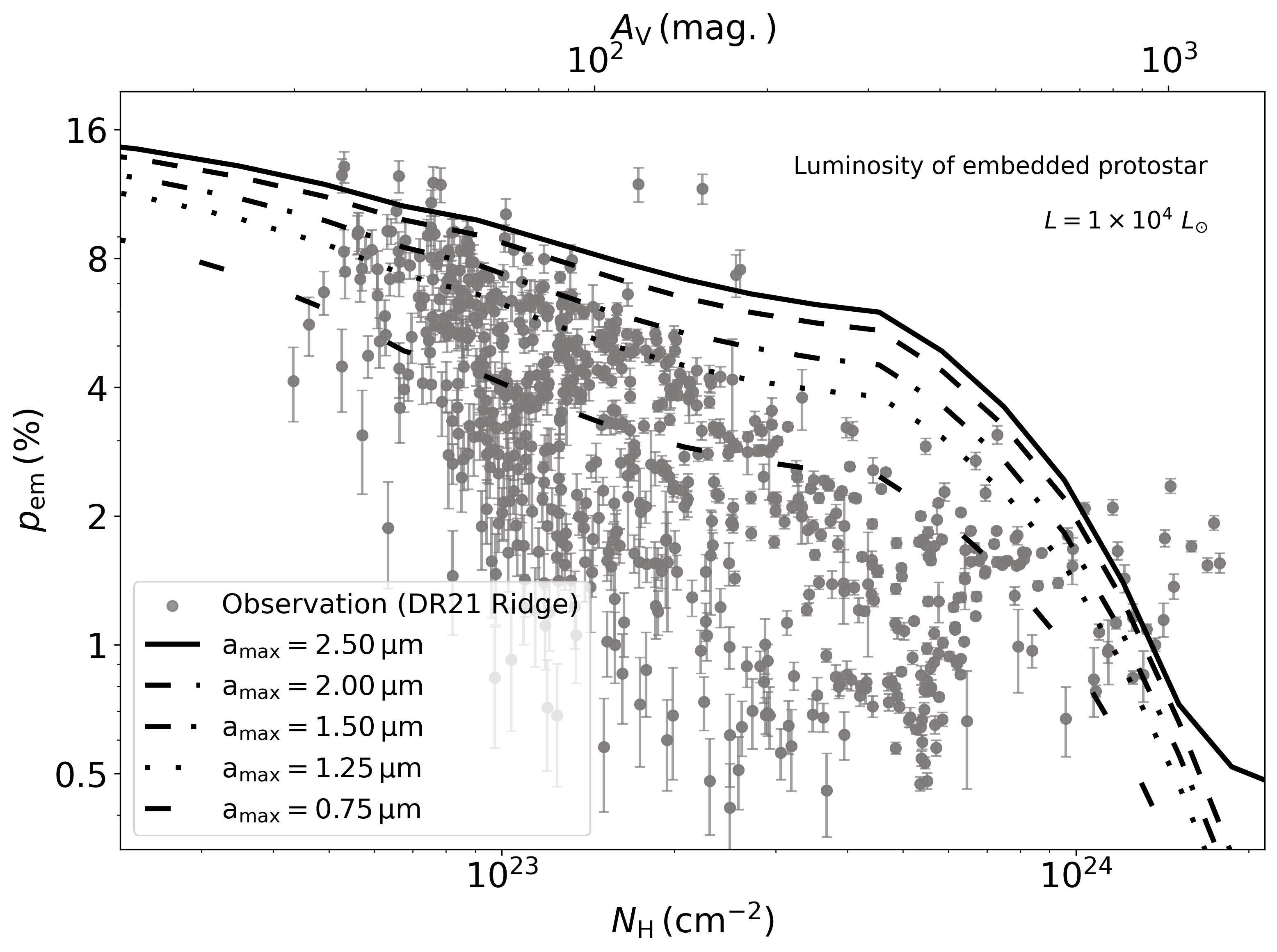}
\caption{
Comparison between radiative transfer model predictions and observed polarization fraction versus hydrogen column density (N$_\mathrm H$) for the DR21 Ridge, for an embedded protostellar luminosity of $1\times10^4 L_{\odot}$. The x-axis on top shows the $A_V$ (mag) extinction values. Each model curve corresponds to a different maximum grain size in the 0.75–2.5 \um{} range.\label{fig:pfracmodel_apendix1}}
\end{figure*}

\begin{figure*}[htb!]
\centering
\includegraphics[scale = 0.5]{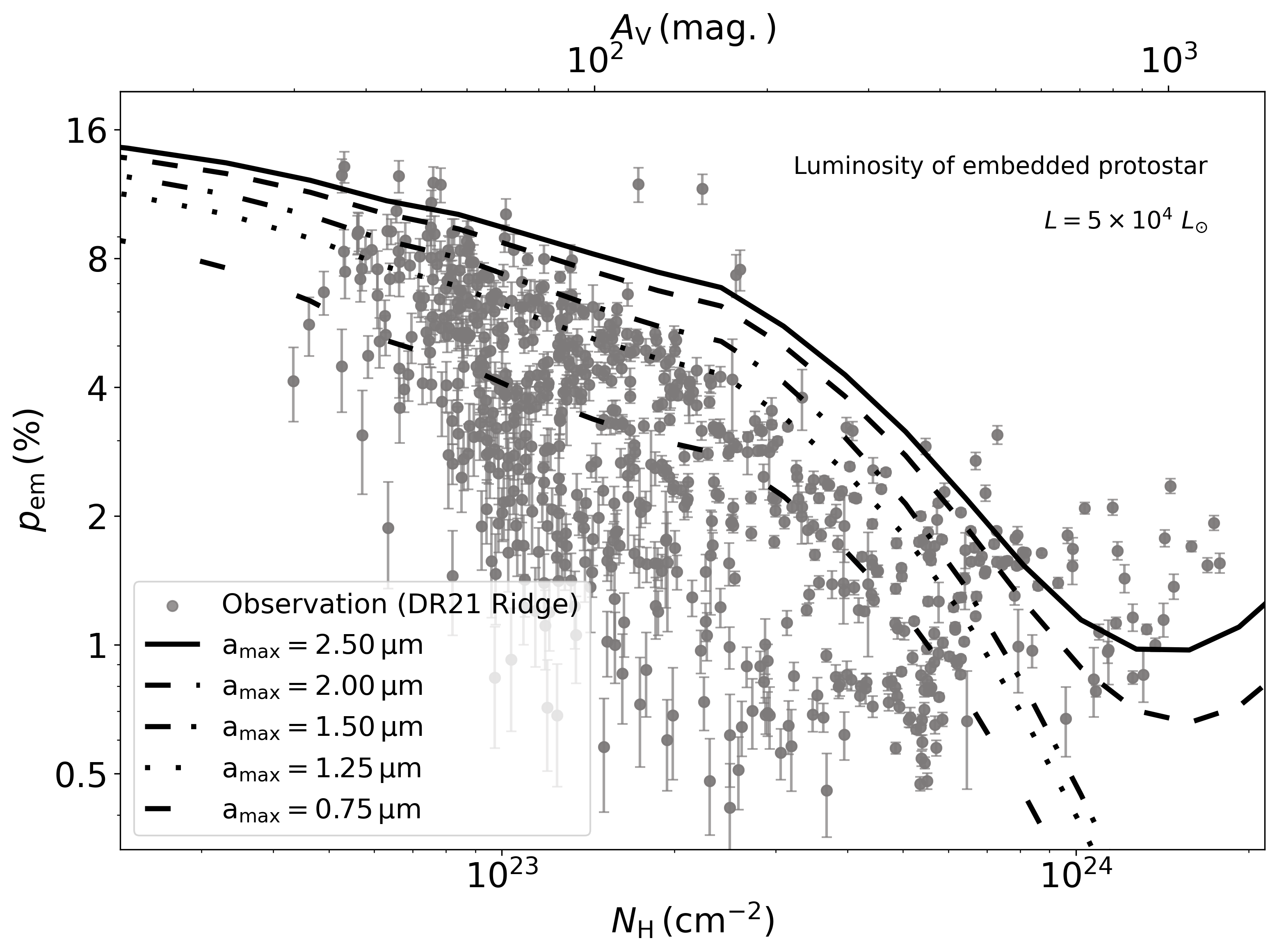}
\caption{Same as Figure \ref{fig:pfracmodel_apendix1}, but assuming an embedded protostellar core luminosity of $5\times10^4 L_{\odot}$. \label{fig:pfracmodel_apendix2}}
\end{figure*}

\begin{figure*}[htb!]
\centering
\includegraphics[scale = 0.5]{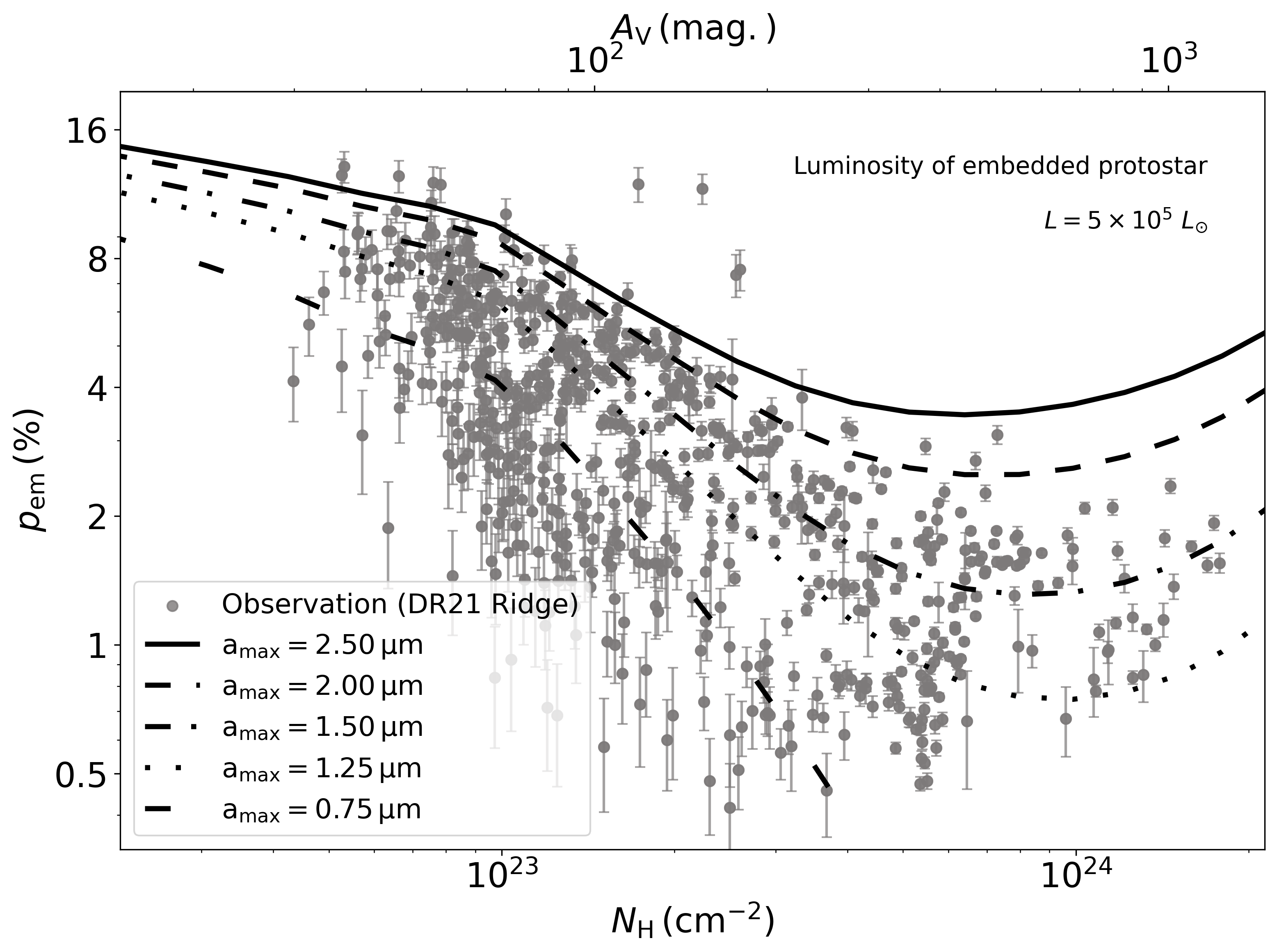}
\caption{Same as Figure \ref{fig:pfracmodel_apendix1}, but assuming an embedded protostellar core luminosity of $5\times10^5 L_{\odot}$. \label{fig:pfracmodel_apendix3}}
\end{figure*}

To complement Section \ref{sec:modeling}, where the embedded protostellar luminosity was set to $1\times10^5~L_{\odot}$, the radiative transfer modeling was repeated for different luminosities. Figures \ref{fig:pfracmodel_apendix1}, \ref{fig:pfracmodel_apendix2}, and \ref{fig:pfracmodel_apendix3} show the modeling results for embedded protostellar luminosities of $1\times10^4~L_{\odot}$, $5\times10^4~L_{\odot}$, and $5\times10^5~L_{\odot}$, respectively. These luminosities were chosen based on the range reported for DR21(OH) in the literature \citep{Harvey1977,Mangum1991,Jakob2007}, which suggests luminosities from $10^{4}~L_{\odot}$ to a few $\times10^{4}~L_{\odot}$, but not exceeding $10^5~L_{\odot}$.

Figure \ref{fig:pfracmodel_apendix1} shows that a luminosity of $1\times10^4~L_{\odot}$ is insufficient to induce RAT reactivation at the highest extinctions. While Figure \ref{fig:pfracmodel_apendix2} shows hints of RAT reactivation for $5\times10^4~L_{\odot}$, the turn-up is predicted only for models with $a_{\mathrm{max}} > 2$~\um{} and at much higher extinctions than observed. Although the literature suggests that the luminosity of DR21(OH) should not exceed $1\times10^5~L_{\odot}$, $L=5\times10^5~L_{\odot}$ was included in the modeling for completeness. The predicted polarization fraction from the $5\times10^5~L_{\odot}$ model agrees well with the observed polarization for $a_{\mathrm{max}} = 1.25$~\um{} and 1.50~\um{}.

\end{appendix}
\end{document}